\documentclass{tlp}
\usepackage{aopmath}

\newtheorem{definition}{Definition}[section]
\newtheorem{example}{Example}[section]
\newtheorem{remark}{Remark}[section]
\usepackage{amssymb}
\usepackage{alltt}
\usepackage[arrow,matrix,frame,curve,cmtip]{xy}\CompileMatrices
\usepackage[english]{babel}
\usepackage{mathrsfs}
\usepackage{pslatex}
\usepackage{url}

\let\from\colon

\let\emptyset\varnothing

\newcommand{\cell}[4]{#1\raisebox{-.8ex}{\mbox{$\stackrel%
{\textstyle\stackrel{#2}{\longrightarrow}}{\scriptstyle #3}$}}#4}

\newcommand{\nat}{\mathbb{N}}
\def\abs#1{\left|#1\right|}
\newcommand{\sbtn}[2]{{#1}/{#2}}
\newcommand{\irule}[2]{\frac{\textstyle\rule[-1.3ex]{0cm}{3ex}#1}%
{\textstyle\rule[-.5ex]{0cm}{3ex}#2}}
\newcommand{\dfun}[3]{#1 \from #2 \to #3}
\newcommand{\eg}{e.g.}
\newcommand{\ie}{i.e.}
\newcommand{\wrt}{w.r.t.}
\newcommand{\cf}{cf.}
\newcommand{\vvdash}{\xymatrix@C=.3pc{{ } \ar@{||-}[r] & { } }}

\newcommand{\Set}{\mathbf{Set}}

\begin{document}
\bibliographystyle{tlp}

\title[Interactive Semantics]
      {An Interactive Semantics of Logic Programming}
\author[R. Bruni, U. Montanari and F. Rossi]
       {ROBERTO BRUNI, UGO MONTANARI\\
        Dipartimento di Informatica, Universit\`a di Pisa,\\
        Corso Italia 40, 56125 Pisa, Italia. \\
        \{\texttt{bruni,ugo}\}\texttt{@di.unipi.it}
       \and
        FRANCESCA ROSSI\\
        Dipartimento di Matematica, Universit\`a di Padova,\\
        Via Belzoni 7, 35131 Padova, Italia. \\
        \texttt{frossi@math.unipd.it}}

\maketitle

\begin{abstract}
 We apply to \emph{logic programming} some recently emerging ideas
 from the field of reduction-based communicating
 systems, with the aim
 of giving evidence of the hidden interactions and the coordination
 mechanisms that rule the operational machinery of such a programming
 paradigm.  The semantic framework we have chosen for presenting our
 results is \emph{tile logic}, which has the advantage of allowing a
 uniform treatment of goals and observations and of applying abstract
 categorical tools for proving the results. As main contributions, we
 mention the finitary presentation of abstract unification, and a
 concurrent and coordinated abstract semantics consistent with the
 most common semantics of logic programming. Moreover, the
 compositionality of the tile semantics is guaranteed by standard
 results, as it reduces to check that the tile systems associated to
 logic programs enjoy the \emph{tile decomposition property}.
 An extension of the approach for handling constraint
 systems is also discussed.
\end{abstract}

\section*{Introduction}

 Logic programming \cite{Llo:FLP} is a foundational research field
 that has been extensively investigated throughout the last 25
 years. It can be said that, in logic programming, theory and practice
 meet together since its very beginning, as each innovation on one
 side contributes many insights to the other side thanks to
 the basic principle of logic programming, which is `writing
 programs by expressing their properties.'
 This symbiosis has also facilitated the study and the prototyping
 of interdisciplinary applications that either extend the `kernel' of
 the framework with additional features or transfer helpful techniques
 from a large variety of paradigms. A typical example is the embedding
 of constraints within logic programming
 \cite{MS:PCI,JM:CLPS}, which retains the declarative and clean semantics of
 logic programming, as well as its typical problem solving features,
 while extending its applicability to many practical
 domains; in fact, constraint logic programming (\textsc{clp}) is now
 considered as a major programming paradigm.

 More interestingly, very often these flows of ideas have been
 profitably bidirectional and continuous, thus allowing one to
 establish strong connections between different areas (bringing useful
 analogies) and also to bridge gaps between different formalisms.

\subsubsection*{Interaction via contextualization and instantiation}

 In this paper, inspired by recent progress in the fields of
 communicating systems and calculi for concurrency, we want to focus
 on an \emph{interactive} view of logic programming. The idea is to
 understand logical predicates as (possibly open) interacting agents whose
 local evolutions are coordinated by the unification engine. In fact,
 the amount of interaction arises from the unification mechanism of
 resolution, as subgoals can share variables and therefore
 `local' progress of a component can influence other components by further
 instantiating such shared variables.
 One central aspect of this view is to understand what kind of
 information we should observe to characterize interaction and how far
 the approach can be extended to deal with different semantic
 interpretations of logic programs.  For example, one interesting
 issue is \emph{compositionality}.  Having a compositional
 semantic framework is indeed very convenient for formal reasoning
 on program properties and can facilitate the development of
 modular programs \cite{BGLM:CSLP,BLM:CMTS,GS:FACS,MP:ALP}.

 We sketch here the main ideas concerning the role played by
 `contexts' in reduction systems, but for a more precise overview we
 invite the interested reader to join us in the little detour, from
 the logic programming world to the process description calculi
 area, inserted in the last part of this introductory
 section (with links to related literature).

 Generally speaking, the issue we focus on is that of equipping a
 reduction system with an interactive semantics. In fact, although
 reduction semantics are often very convenient because of a friendly
 presentation, they are not compositional `in principle.'  The problem
 is that they are designed having in mind a progressive reduction of
 the initial state to a suitable normal form, \ie, one focuses
 on a completely specified system that can be studied in
 isolation from all the rest. In logic programming, this would
 correspond to studying the refutation of ground goals only and to
 develop an \emph{ad-hoc} system to this aim.  Then, if one wants to
 study the semantics of partially specified components the framework
 is no longer adequate and some extensions become necessary. For
 example, in process description calculi, a partially specified
 component can be a process term (called \emph{open process} or
 \emph{context}) that contains suitable process variables representing
 generic subprocesses. However, also a closed term (\ie, without free
 process variables) can be considered an open system when it evolves
 as part of a broader system, by
 interacting with the environment.  In logic
 programming we can distinguish two main kinds of openness and
 interaction. A first kind is due to goals with variables (rather than
 ground) that can obviously be regarded as partially specified
 systems. A second kind consists of regarding an atomic
 goal as part of a larger conjoined goal with which it must interact.

 The obvious way to deal with partially specified components is to
 transform the problem into the reduction case, which we know how to
 solve. This means that (1) the variables of open
 processes will be instantiated in all possible ways to obtain closed
 systems that can be studied; (2) in order to study the semantics of
 closed subprocesses we will insert them in all possible
 contexts and then study their reductions.  Moreover, the
 operations of contextual and instantiation closure can be rendered
 dynamically, provided that one defines a \emph{labeled transition
 system} (\textsc{lts}) whose labels record the information on the
 performed closure, and this has originated the idea of observing
 contexts and instantiations (sometimes called \emph{external} and
 \emph{internal} contexts, respectively).

 Even if these views can look semantically adequate, it can
 be noted, as their main drawback, that they are not applicable in
 practice, because all considered closures are infinitary.
 The situation can be improved if one is able to identify a small
 finite set of contexts and/or instances that contains all useful
 information, since this can make the approach operationally
 satisfactory. While a general methodology for accomplishing this task
 in  communicating and mobile calculi is difficult to find
 (see \eg, \cite{LM:DBCRS}), we think that logic programming
 represents the perfect situation where it is possible to fully develop
 the closure approach.

 When dealing with the interactive view of logic programs, the idea is
 that unification is the basic action taking place during
 computation, and therefore the observed information must rely on such
 an action. We have seen that two kinds of closure can be
 distinguished that are dual to each other, namely
 \emph{contextualization} and \emph{instantiation}. The former can be
 used to embed components in a larger environment, while the latter
 serves to specialize an open system to some particular instance.

 We shall concentrate our efforts on \emph{pure} logic
 programming (\ie, classical Horn clauses, without any additional
 `gadgets'). Hence contextualization corresponds to putting the goal
 in conjunction with other goals,\footnote{In pure logic programming,
 contextualization does not provide any
 additional information on the possible reductions, as the head
 of each clause consists of only one predicate. The situation
 would be different if generalized multi-head Horn clauses were
 considered, a topic that will be discussed in the conclusions, or
 if second order logic were considered (other predicate contexts
 should be considered beside conjunction).}
 \ie, given a goal $G$ we should
 put it in the context $\_\wedge G'$ for all possible $G'$.
 With respect to instantiation, our proposal is to regard the computed
 substitutions for the variables in the (sub)goals as observable
 internal contexts, which further instantiate the system
 components. Thus, given a goal $G$, we can apply the
 substitution $\sigma$ to the free variables of $G$ and study the
 consequent changes in the semantics.

 The analogy and distinction between internal and external contexts
 become clear if we look at the term algebra over a signature
 $\Sigma$ from a categorical perspective: the objects of the category
 are underlined natural numbers, an $n$-tuple of terms over $m$
 variables corresponds to an arrow from $\underline{m}$ to
 $\underline{n}$, and composition of arrows $t_1\from \underline{m}
 \to \underline{k}$ and $t_2 \from \underline{k} \to \underline{n}$
 is given by substituting the $k$ variables in $t_2$ by the corresponding
 terms in the tuple $t_1$. Then, composing to the right means
 inserting in a context, while composing to the left means providing
 an internal context (\eg, $t_2$ above is external to $t_1$, while
 $t_1$ is internal to $t_2$).

\subsubsection*{Tile logic as a semantic framework}

 For pursuing this research programme, we have chosen to rely on
 \emph{tile logic} \cite{GM:RRCCS,GM:TM} that can provide a convenient
 abstract computational model for logic programming, where many of the
 discussed aspects can be suitably represented and managed.

 The tile framework takes inspiration from and bears many analogies
 with various \textsc{sos} formats
 \cite{Plo:SAOS,DeS:HLSD,BIM:BCBT,GV:SOSBC,Ber:CTSOS},
 \emph{context systems} \cite{LX:CTOSC},
 \emph{structured transition systems} \cite{CM:ASSTS},
 and \emph{rewriting logic} \cite{Mes:CRL}. It allows to define models
 that are compositional both in `space' (\ie, according to the
 structure of the system) and in `time' (\ie, according to the
 computation flow). In particular, tile logic extends rewriting logic
 with a built-in mechanism, based on observable effects, for
 coordinating local rewrites. The effects are in some sense the
 counterparts of labels in \textsc{lts} operational semantics.
 However, since tiles
 are designed for dealing with open states (as opposed to the ordinary
 `ground' view of \textsc{lts}'s generated from \textsc{sos} rules),
 they seem more apt for many
 applications. The idea is to employ a set of rules (called
 \emph{tiles}) to define the behavior of partially specified
 components (\ie, components that can contain variables), called
 \emph{configurations}, only in terms of the possible interactions
 with the internal/external environment.
 In this way, the behavior of a system must be
 described as a coordinated evolution of its local
 subconfigurations. The name `tile' is due to the graphical appearance
 of such rules, which have the form in Figure~\ref{atilefig}, also
 written $\alpha:\cell{t}{u}{v}{s}$, stating that the \emph{initial
 configuration} $t$ evolves to the \emph{final configuration} $s$ via
 the tile $\alpha$, producing the \emph{effect} $v$, which can be
 observed by the rest of the system, but such a step is allowed only
 if the subcomponents of $t$ (\ie, the arguments to which $t$ is
 connected via its input interface) evolve to the subcomponents of
 $s$, producing the effect $u$, which acts as the \emph{trigger} for
 the application of $\alpha$. Triggers and effects are called
 \emph{observations} and tile vertices are called
 \emph{interfaces}. The arrows $t$, $u$, $v$ and $s$ form the
 \emph{border} of $\alpha$.

\begin{figure}[t]
\begin{center}
$\xymatrix@C=6pc@R+1pc{
 *[o]=<.5pc>{\ }\drop\cir{}
 \ar[r]^{t}
 \ar[d]_{u}
 \ar@{}[rd]|{\alpha}
 \ar@<2.5pc>@{}[r]|{\mathrm{initial}}
 \ar@<1.5pc>@{}[r]|{\mathrm{configuration}}
 \ar@<-2.5pc>@{}[d]|{\mbox{trigger}}
 \POS[]+<-2pc,2.5pc>\drop{\mathit{initial}}
 \POS[]+<-2pc,1.5pc>\drop{\mathit{input}}
 \POS[]+<-2pc,.5pc>\drop{\mathit{interface}}
 &
 *[o]=<.5pc>{\ }\drop\cir{}
 \ar[d]^{v}
 \ar@<2.5pc>@{}[d]|{\mbox{effect}}
 \POS[]+<2pc,2.5pc>\drop{\mathit{initial}}
 \POS[]+<2pc,1.5pc>\drop{\mathit{output}}
 \POS[]+<2pc,.5pc>\drop{\mathit{interface}}
 \\
 *[o]=<.5pc>{\ }\drop\cir{}
 \ar[r]_{s}
 \ar@<-1.5pc>@{}[r]|{\mathrm{final}}
 \ar@<-2.5pc>@{}[r]|{\mathrm{configuration}}
 \POS[]-<2pc,.5pc>\drop{\mathit{final}}
 \POS[]-<2pc,1.5pc>\drop{\mathit{input}}
 \POS[]-<2pc,2.5pc>\drop{\mathit{interface}}
 &
 *[o]=<.5pc>{\ }\drop\cir{}
 \POS[]+<2pc,-.5pc>\drop{\mathit{final}}
 \POS[]+<2pc,-1.5pc>\drop{\mathit{output}}
 \POS[]+<2pc,-2.5pc>\drop{\mathit{interface}}
}$
\end{center}
\caption{A tile.}
\protect\label{atilefig}
\end{figure}

\begin{figure}[t]
\begin{center}
$\vcenter{\xymatrix@R-1pc@C+1pc{
 {\circ} \ar[r] \ar[d] \ar@{}[rd]|{\alpha} &
 {\circ} \ar[r] \ar[d] \ar@{}[rd]|{\beta} &
 {\circ} \ar[d] \\
 {\circ} \ar[r] &
 {\circ} \ar[r] &
 {\circ} }}$
\hfill $\vcenter{\xymatrix@R-1pc@C+1pc{
 {\circ} \ar[r] \ar[d] \ar@{}[rd]|{\alpha} &
 {\circ} \ar[d] \\
 {\circ} \ar[r] \ar[d] \ar@{}[rd]|{\beta} &
 {\circ} \ar[d] \\
 {\circ} \ar[r] &
 {\circ} }}$
\hfill $\vcenter{\xymatrix@R-2pc@C-1pc{
 & {\circ} \ar[rr] \ar[dd] & &
 {\circ} \ar[dd] \\
 {\circ} \ar[rr] \ar[dd] & &
 {\circ} \ar[dd] \ar@{}[ru]|{\beta} \\ &
 {\circ} \ar[rr] & & {\circ} \\
 {\circ} \ar[rr] \ar@{}[ru]|{\alpha} &
 {\ } &
 {\circ} }}$
\end{center}
\caption{Horizontal, vertical and parallel tile
compositions.} \protect\label{threecomptile}
\end{figure}
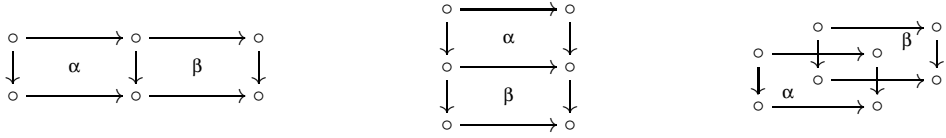

 Tiles can be composed horizontally, vertically, and in parallel
 to generate larger steps. The three compositions are
 illustrated in Figure~\ref{threecomptile}. Horizontal composition
 yields rewriting synchronization (\eg, between the evolution of
 an argument via $\alpha$ and the evolution of its
 environment via $\beta$, as the effect of $\alpha$
 provides the trigger for $\beta$). Vertical
 composition models the sequential
 composition of computations.  The operation of parallel composition
 corresponds to building concurrent steps, where two (or more)
 disjoint configurations can concurrently evolve. Of course, the
 border of a concurrent step is the parallel composition of the
 borders of each component of the step.

 Given a set of basic tiles, the associated \emph{tile logic} is
 obtained by adding some canonical `auxiliary' tiles and then closing
 by (the three kinds of) composition both
 auxiliary and basic tiles. As an example, auxiliary tiles may be
 introduced that accommodate isomorphic transformations of interfaces,
 yielding consistent rearrangements of configurations and observations
 \cite{BMM:PTTL,Bru:TL}.

 Tile logic deals with algebraic structures on configurations that can
 be different from the ordinary, tree-like presentation of terms
 employed in most \textsc{lts}'s. All these structures, ranging from
 graphs and term graphs to partitions and relations, give rise to
 monoidal categories and, therefore, possess the two basic operations
 needed by tile configurations. This is very convenient, as the models
 of the logic can be formulated in terms of \emph{monoidal double
 categories}. In this paper, we assume the reader to be familiar with
 the basic concepts of category theory, though we shall not push their
 usage too far (employing categories in a mild way) and shall give
 informal explanations of most categorical constructs introduced.

 Likewise context systems \cite{LX:CTOSC} and \emph{conditional
 transition systems} \cite{Ren:BOT}, tile logic allows one to reason
 about terms with variables \cite{BFMM:BC}. This means, \eg, that
 trace semantics and bisimilarity can be extended straightforwardly to
 open terms by taking as observation the pair $\langle
 \mathit{trigger}, \mathit{effect}\rangle$, whereas ordinary
 \textsc{lts}'s deal with transitions from closed terms to closed
 terms for which triggers are trivial identities. The compositionality
 of abstract semantics (either based on traces or on bisimilarity) can
 then be guaranteed by algebraic properties of the tile system or by
 suitable specification formats \cite{BFMM:BC}. In particular, we
 shall see that the \emph{decomposition property} \cite{GM:TM} yields
 a very simple proof of the compositionality of the tile logic
 associated to a logic program.

\subsubsection*{The tile approach to logic programming}

 A well-known fact (\cf\ the discussion in Section~\ref{RVPsec}
 and \cite{BR:CUA,CM:ASSTS}) that is exploited in the construction we
 propose is that in categorical terms the construction of the
 \emph{most general unifier} (mgu) between a subgoal and the head
 of a clause can be expressed as a \emph{pullback} in the
 syntactic category associated to the signature under
 consideration. One of the contributions of this paper is in fact
 to provide a constructive, modular way of building the pullback
 construction. It is similar to the ordinary unification mechanism
 but formulated in a completely abstract way by means of
 coordination rules. This translates immediately in terms of tile
 logic, completing the first part of our research programme, that
 is, understanding the extent of interaction we shall observe,
 and expressing it in a formal system.

 For the rest, we define a transformation from logic programs to
 tile systems by associating a basic tile to each Horn clause in the
 program.  Then, the resulting tile models are shown to provide a
 computational and semantic framework where the program and its
 tile representation yield exactly the same set of computed answer
 substitutions for any goal. It is remarkable that all
 aspects concerning the control flow are now automatically
 handled by tiles (e.g. generation of fresh variables, unification,
 construction of resolvents).

 One of the advantages of tile logic is to make evident the duality
 between contextualization and instantiation, still being able to deal
 in a uniform way with both perspectives. The same thing can be said
 for the uniform treatment of configurations and observations that
 facilitates the use of contexts as labels, providing many insights on
 the way the basic pieces can be put together to form a whole
 computation.

 The tile presentation of a program allows us not only to
 transfer to logic programming abstract semantic equivalences
 based on traces and bisimilarity, but also to show that
 these equivalences are compositional (\ie, they are
 \emph{congruences}) via an abstract proof based on the
 decomposition property of the underlying tile system. More
 concretely, denoting by $\sim$ any of the two equivalences (on
 goals) mentioned above, we have almost for free that if
 $G_1\mathrel{\sim} G_2$, then: (1) $\sigma(G_1) \mathrel{\sim}
 \sigma(G_2)$ for any substitution $\sigma$, and (2) $G_1\wedge G
 \mathrel{\sim} G_2 \wedge G$ for any goal $G$.
 (This lifts also to the case where the simpler `success' semantics
 is considered.)

 The application of our `tile' techniques to logic programming can
 serve as a basis for establishing useful connections and studying
 analogies with the process calculi paradigm. For example,
 it comes out that there is a strong resemblance between the
 parallel operator of many process calculi and the conjunction
 operator on goals. As another example, it would be interesting to
 transfer to logic programming concepts like `explicit substitution'
 and `term graph', which play important roles in the implementation
 of distributed systems.

\subsubsection*{A digression: Sources of inspiration}

 Before illustrating the organization of the material, we want to
 explain more precisely the intuition that motivated our research on
 interactive semantics for reduction systems and its application to
 logic programming.
 As already pointed out, our sources of inspiration mostly come
 from contributions in the theory of communicating systems.  The
 first fact to note is that there are two well recognized and
 widely studied schools of thought for giving semantics to
 process description calculi, namely via \emph{reduction} rules (especially
 popular after Berry and Boudol's \textsc{cham} \cite{BB:CAM})
 and via \textsc{lts}'s.

 The first approach relies on the assumption that it is possible to
 observe and manipulate the global state of a complex system. In particular,
 the current state can be inspected for finding a
 \emph{redex}, \ie, a candidate for the application of a reduction step. The
 redexes usually coordinate the activity of several logically distinct
 components of the system, and therefore, to some extent, the
 reduction step synchronizes their local activities into a global
 atomic move. For dealing with compositionality, one would
 be interested in deriving the semantics of a whole entity in terms of
 the semantics of its very basic component parts, which can become a
 hard task when reductions are global actions.  The
 problem is that a redex may lie in between a component and the
 external environment, and therefore to understand the behavior of a
 component as a stand alone entity, we have to consider its
 interactions with all possible environments, in the style of
 \emph{testing semantics} \cite{dNH:TEP}.

 Instead, the point of view of observational equivalences based on
 \textsc{lts} semantics is to use \emph{observations} (transition labels)
 to derive observational equivalences on processes,
 as \eg, \emph{bisimilarity} \cite{Par:CAIS,Mil:CCS}. Moreover, the
 formats for specifying \textsc{lts} operational semantics can exploit
 inductively the structure of a complex state to define its semantics
 in terms of the actions that can be accomplished by
 subcomponents, guaranteeing compositionality properties
 like `bisimilarity is a congruence.'

 One emerging idea to provide reduction semantics with an
 interactive, observational view is that of `observing contexts'
 \cite{Mil:PPC,MS:DCPB,Mil:CFI,Ber:CTSOS,Sew:RRBC,LM:DBCRS,CLM:CECAG}.
 Basically, starting from a reduction system, one has to define
 the semantics of a local component by embedding it in all
 possible contexts and by considering those contexts as
 observations. Then, when several components are assembled
 together, it is possible to predict the semantics of the result
 simply by inspecting the behaviors of each component in the
 environment contributed by all the remaining components.  This
 approach gives rise to a special kind of bisimulation, called
 \emph{dynamic bisimilarity} \cite{MS:DCPB}, which is the
 coarsest congruence that is also an ordinary bisimulation.
 This approach
 corresponds to some extent to give the possibility to
 dynamically reconfigure the system and has also some
 applications to open ended systems \cite{BMS:OES}.
 Though theoretically sound, this solution
 leaves open many operational questions, because the semantics
 must take into account all possible contexts.

 Many people attempt to define a general and clever
 methodology for passing from reduction semantics to
 (compositional) \textsc{lts} semantics
 \cite{Sew:RRBC,LM:DBCRS,CLM:CECAG}. In particular, Leifer and
 Milner show in \cite{LM:DBCRS} that a minimal set of contexts
 is definable whenever sufficiently many \emph{relative pushouts}
 exist in the category of configurations. Roughly speaking, it
 must be the case that for any configuration $t$ and any reduction
 rule with which $t$ can react in a suitable
 environment $C$, then there exists a minimal observable context
 $C'$ that makes such reduction possible.

 Dual to the problem of `contextualization' is the problem of
 `instantiation.' It arises when one wants to extend the
 compositionality from ground processes to open processes.
 In fact, the equivalence on open terms is usually defined
 via the equivalence on closed terms, by saying that two
 contexts are equivalent if their closures
 under all possible ground instantiations are so. Again, it is
 preferable to avoid the instantiation closure and find a more compact
 way to enforce the modularity of the framework. This issue has been
 pursued in two recent works \cite{Ren:BOT,BFMM:BC} for providing
 general specification formats that guarantee the compositionality of
 open systems. They are based on the idea of recording in the
 transition labels not only the effects of each move, but also the
 triggers provided by the subcomponents for applying the transition to
 the global state. Consequently, in the `dynamic' version, instantiation
 becomes a sort of `internal contextualization' and
 substitutions can be used as labels (in the trigger part).

 In the case of logic programming, many of the above concepts find a
 natural meaning. Thus, \eg, goal instantiation
 is a relevant internal contextualization that can
 modify the semantics of the goal (\eg, by making impossible the
 unification with the head of a clause which can otherwise be
 applicable), while external contextualization is given by conjunction with
 other goals (it can be relevant when multi-headed
 clauses are allowed).

\subsubsection*{Structure of the paper}

 We fix the notation and recall the necessary
 background in Section~\ref{Bsec}. Due to the heterogeneity of the
 material, its presentation is separated in four
 parts: Section~\ref{Nsec} summarizes a few elementary
 definitions about signatures, substitutions and Horn clauses;
 Section~\ref{OSsec} recalls the operational machinery of logic
 programming; Section~\ref{TLsec} presents the tile notation and
 the categorical models based on double categories;
 Section~\ref{ATsec} presents the concepts of Section~\ref{Nsec}
 under a different light (exploiting Lawvere's pioneering work
 \cite{Law:FSAT}), which will offer a more convenient notation
 for representing logic programs in tile logic. While the
 contents of Sections~\ref{Nsec} and~\ref{OSsec} are standard,
 the notions recalled in Sections~\ref{TLsec} and~\ref{ATsec}
 might be not so familiar to the logic programming community.

 In Section~\ref{RVPsec} we recall the ways in which
 most general unifiers, equalizers and pullbacks intertwine.
 This should provide the
 reader with the formal knowledge for understanding the technical
 details of the correspondence between unification in logic
 programming and coordination via pullback tiles, which is
 explained in Section~\ref{DCPsec}. In particular, we think that
 the results in Section~\ref{FPPsec} are the key to the
 application of tiles to logic programming.

 Section~\ref{TSLPsec} exploits the notation and results from
 Section~\ref{DCPsec} to establish the connection between logic
 programming and tile logic. The transformation is described in
 Section~\ref{FLPTLTsec}, together with a simple example that
 illustrates the correspondence between the two views. The main
 advantages of the tile approach are examined separately in
 Section~\ref{FCTTsec} (connections with ongoing research on process
 calculi), Section~\ref{ROSsec} (formal correspondence with ordinary
 semantics), Section~\ref{GCsec} (goal compositionality via abstract
 congruences), Section~\ref{THREEsec} (comparison between
 goal equivalences obtained by considering different instantiation
 closures), and in Section~\ref{DTsec} (insights on concurrency and
 coordination). The compositionality of the resulting framework
 strongly depends on the representation results of
 Section~\ref{FPPsec}, that allow one to decompose a complex
 coordination along its basic bits.

 While the paper focuses on pure logic programs, we think that the
 approach can be extended to take into account many variants of logic
 programming. Some of these extensions are discussed in
 Section~\ref{CLPsec} (devoted to constraint logic programming)
 and in the concluding section.

\section{Background}\label{Bsec}

\subsection{Notation}\label{Nsec}

 Let $\Sigma$ be a two sorted signature over the set of sorts
 $\{\mathsf{t},\mathsf{p}\}$. Provided that
 the sort $\mathsf{p}$ does not appear in the arity of any
 operator, we call $\Sigma$ a \emph{logic
 program signature} and we denote by $\Sigma_\Phi =
 \bigcup_n\Sigma^n_\Phi$ and $\Sigma_\Pi = \bigcup_n\Sigma^n_\Pi$
 the ranked sets of \emph{function symbols} $\dfun
 f{\mathsf{t}^n}{\mathsf{t}}$ and of \emph{predicate symbols}
 $\dfun p{\mathsf{t}^n}{\mathsf{p}}$, respectively.

 As usual, given a set $X$ of (term) variables, we denote with
 $\mathbb{T}_\Sigma(X)$ the free $\Sigma$-algebra generated by $X$.  A {\em
 term over} $X$ is an element of $\mathbb{T}_{\Sigma_\Phi}(X)$. The set of all
 \emph{ground terms} (\ie, terms without variables) is called the {\em
 Herbrand universe for} $\Sigma$. An \emph{atomic formula over} $X$ has
 the form $p(t_1,\ldots,t_n)$ where $p\in\Sigma^n_\Pi$ and
 $t_1,\ldots,t_n$ are terms over $X$. A \emph{conjunctive formula} is
 just a tuple of atomic formulas. The set of all ground atomic
 formulas is called the \emph{Herbrand base} for $\Sigma$.

 If $X = \{x_1,\ldots,x_n\}$ and $Y$ are sets of variables, a {\em
 substitution from $Y$ to $X$} is a function
 $\dfun\sigma{X}{\mathbb{T}_{\Sigma_\Phi}(Y)}$, usually denoted by
 $[\sbtn{\sigma(x_1)}{x_1},\ldots,\sbtn{\sigma(x_n)}{x_n}]$.  If $t$
 is a term over $X$ and $\sigma$ is a substitution from $Y$
 to $X$ then the term over $Y$ obtained by {\em
 simultaneously} substituting in $t$ all the occurrences of the
 variables in $X$ with their images through $\sigma$ is called the {\em
 application of $\sigma$ to $t$} and written $\sigma;t$.

 If $\sigma$ is a substitution from $Y$ to $X$ and $\sigma'$ is a
 substitution from $Z$ to $Y$, their \emph{composition} is the
 substitution $\sigma';\sigma$ from $Z$ to $X$ defined by applying
 $\sigma'$ to each image of the variables in $X$ through $\sigma$.  A
 substitution $\sigma$ is said to be \emph{more general} than $\sigma'$
 if there exists a substitution $\theta$ such that $\sigma' = \theta;\sigma$.
 It is worth noticing that since substitution composition is
 associative with the identity substitutions
 $[\sbtn{x_1}{x_1},\ldots,\sbtn{x_n}{x_n}]$ as neutral elements, then
 substitutions form the arrows of a category having finite sets of
 variables as objects.

 Two terms (also atomic formulas) $t$ and $s$ \emph{unify} if there
 exists a substitution $\theta$ such that $\theta;t = \theta;s$. In
 this case $\theta$ is called a \emph{unifier} of $t$ and $s$. If $t$
 and $s$ unify there exists also a \emph{most general unifier} (unique
 up to variable renaming), \emph{mgu} for short.

 The mgu can be computed by employing, \eg, the following
 (nondeterministic) algorithm that operates on a set of equations (at
 the beginning the set is the singleton $\{t=s\}$).

{\em
\begin{itemize}
\item
 Apply one of the following steps until stability is reached:
 \begin{enumerate}
 \item
  eliminate the equation $x=x$ from the set for some variable $x$;
 \item
  eliminate the equation $f(t_1,\ldots,t_n) = f(s_1,\ldots,s_n)$ from
  the set and insert the equations $t_1=s_1$,\ldots,$t_n=s_n$ in it;
 \item
  if the equation $x=t$ with $x$ a variable not appearing in $t$ is
  contained in the current set, apply the substitution $[\sbtn{t}{x}]$
  to all the other equations (but do not remove $x=t$).
  \end{enumerate}
\end{itemize}
}

 The algorithm always terminates. It terminates with success if
 the resulting set of equations has the form $\{x_1 =
 t_1,\ldots,x_n=t_n\}$ with $x_i$ not appearing in $t_j$ for all
 $i,j\in [1,n]$. The algorithm can be efficiently computed if it
 cyclically executes the sequence of steps
 1, 3 (with $x=y$), 2, 3 (with $x=t$).

\subsection{Syntax and operational semantics of logic programs}\label{OSsec}

 In this section we briefly recall the basics of the operational
 semantics of logic programs. We refer to \cite{Llo:FLP} for a more
 detailed introduction to the subject.

 A \emph{definite Horn clause} $c$ is an expression of the form

\[H \mathrel{:-} B_1,\ldots,B_n\]

\noindent
 with $n\geq 0$, where $H$ is an atomic formula called
 the \emph{head} of $c$ and $\langle B_1,\ldots,B_n\rangle$ is a
 (conjunctive) formula called the \emph{body} of $c$.
 A logic program $\mathcal{P}$ is a finite collection of clauses
 $\{c_1,\ldots,c_m\}$.

 A \emph{goal} is an expression of the form

\[\mathrel{?-} A_1,\ldots,A_k\]

\noindent
 with $k\geq 0$, where $G \mathrel{\equiv} \langle
 A_1,\ldots,A_k\rangle$ is a (conjunctive) formula and the $A_i$'s are
 the \emph{atomic subgoals} of $G$. If $k=0$, then $G$ is called the
 \emph{empty goal} and is denoted by `$\Box$.'

 Given a goal $G \mathrel{\equiv} \langle A_1,\ldots,A_k\rangle$ and a
 clause $c\equiv H \mathrel{:-} B_1,\ldots,B_n$ (with all variables in the
 latter possibly renamed to avoid confusion with those in $G$) an
 \emph{(SLD)-resolution step} involves the selection of an atomic goal
 $A_i$ such that $H$ and $A_i$ unify and the construction of their mgu
 $\theta$. The step leads to a new goal

\begin{eqnarray*}
 G'
 & \mathrel{\equiv} &
 \theta;\langle A_1,\ldots,A_{i-1},
                B_1,\ldots,B_n,
                A_{i+1},\ldots,A_k\rangle \\
 & \mathrel{\equiv} &
                \theta;A_1,\ldots,\theta;A_{i-1},
                \theta;B_1,\ldots,\theta;B_n,
                \theta;A_{i+1},\ldots,\theta;A_k
\end{eqnarray*}

\noindent
 which is called the \emph{resolvent} of $G$ and $c$. In this case we
 say that $G'$ is \emph{derived from $G$ and $c$ via $\theta$} and we
 write $G\Rightarrow_{c,\theta} G'$ or simply $G\Rightarrow_{\theta}
 G'$.

 Given a logic program $\mathcal{P} = \{c_1,\ldots,c_m\}$ and a goal
 $G_0$, an {\em (SLD-)derivation} of $G_0$ in $\mathcal{P}$ is a
 (finite or infinite) sequence $G^0,G^1,G^2,\ldots$ of goals, a
 sequence $c_{i_1},c_{i_2},\ldots$ of (renamed) clauses and a sequence
 $\theta_1,\theta_2,\ldots$ of mgu's such that $G^{i+1}$ is derived
 from $G^i$ and $c_{j_i}$ via $\theta_{i}$.  An
 \emph{(SLD-)refutation} of $G_0$ is a finite derivation of $G_0$
 ending with the empty goal. In this case the substitution $\theta =
 (\theta_l;\cdots;\theta_1)_{\mid Var(G)}$ is called a {\em computed
 answer substitution} for $G$, written $G\Rightarrow^*_\theta
 \Box$. The `small-step' operational semantics is formalized in
 Table~\ref{smcasruletab} (but in the rule for atomic goal we must be
 certain that $\rho$ renames the variables in the clause by globally
 fresh names).

\begin{table}
\caption{Operational rules for SLD-resolution (small step
semantics).} \protect\label{smcasruletab}
\begin{tabular}{ccl}
\hline \hline
 $\irule{}
        {\mathcal{P}\vvdash \Box,G\Rightarrow_{id} G}\ \ \
  \irule{}
        {\mathcal{P}\vvdash G,\Box\Rightarrow_{id} G}$ & & empty goal \\
\\
 $\irule{(H \mathrel{:-} F) \in \mathcal{P}\ \
         \sigma = \mathrm{mgu}(A,\rho;H)}
        {\mathcal{P}\vvdash A\Rightarrow_{\sigma}
                            \sigma;\rho;F}$ & & atomic goal \\
\\
 $\irule{\mathcal{P}\vvdash G\Rightarrow_{\sigma}F}
        {\mathcal{P}\vvdash G,G'\Rightarrow_{\sigma}
                            F,(\sigma;G')}\ \ \
  \irule{\mathcal{P}\vvdash G\Rightarrow_{\sigma}F}
        {\mathcal{P}\vvdash G',G\Rightarrow_{\sigma}
                            (\sigma;G'),F}$ & & conjunctive goal \\
\hline \hline
\end{tabular}
\end{table}

 The inductive definition of computed answer substitution and
 refutation in the `big-step' style is given by the rules in
 Table~\ref{casruletab}. The notation $\mathcal{P}\vdash_{\theta} G$
 means that $\mathcal{P}\vvdash G \Rightarrow^*_\theta \Box$, \ie,
 that the goal $G$ can be refuted by using clauses in the program
 $\mathcal{P}$. The first rule says that the empty goal can always be
 refuted with the empty computed answer substitution $\epsilon$. The
 second clause says that an atomic goal can be refuted provided that
 it can be unified with the head $H$ of a clause (suitably renamed by
 $\rho$ to avoid name conflicts with $A$) in the program $\mathcal{P}$
 via the mgu $\sigma$ and that the goal obtained by applying $\sigma$
 to the (renamed) body $F$ of the clause can be refuted with $\theta$.
 The third rule says that a conjunctive goal can be refuted provided
 that its leftmost subgoal can be refuted first with $\sigma$, and
 then the goal obtained by applying $\sigma$ to the other subgoals can
 be refuted with $\theta$. Although imposing a sequentialization in
 the resolution process can appear as an arbitrary choice, the fact
 that refutation involves finite derivations and the well known
 switching lemma guarantee the completeness of the formal system.

\begin{table}
\caption{Operational rules for SLD-resolution (big step
semantics).} \protect\label{casruletab}
\begin{tabular}{rcl}
\hline \hline
 $\irule{}{\mathcal{P}\vdash_{\epsilon} \Box}$ & & empty goal \\
\\
 $\irule{(H \mathrel{:-} F) \in \mathcal{P}\ \
         \sigma = \mathrm{mgu}(A,\rho;H)\ \
         \mathcal{P}\vdash_{\theta} \sigma;\rho;F}
        {\mathcal{P}\vdash_{\theta;\sigma} A}$ & & atomic goal \\
\\
 $\irule{\mathcal{P}\vdash_{\sigma} A\ \
         \mathcal{P}\vdash_{\theta} \sigma;F}
        {\mathcal{P}\vdash_{\theta;\sigma} A,F}$ & & conjunctive goal \\
\hline \hline
\end{tabular}
\end{table}

\subsection{Double categories and tile logic}\label{TLsec}

 The point of view of tile logic \cite{GM:TM,Bru:TL} is that the
 dynamics of a complex system can be better understood if we reason in
 terms of its basic components and of the interactions between
 them. Therefore, reductions must carry observable information about
 the triggers and the effects of the local step. This extends the
 point of view of rewriting logic, where reductions can be freely
 nested inside any context (and also freely instantiated): In tile
 logic, contextualization and instantiation are subordinated to the
 synchronization of the arguments with the environment, \ie, the
 effect of the tile defining the evolution of the former must provide
 the trigger for the evolution of the second. When the coordination is
 not possible, then the step cannot be performed.

 An abstract account of the connections between states and
 dynamics can be given via (monoidal) \emph{double
 categories}, by exploiting their two-fold representation: one
 dimension is for composing states and the second dimension is
 for composing computations. In fact, the models of tile logic are
 suitable double categories \cite{GM:TM,BMM:SMCDC}, and the basic
 tiles of a tile system provide a finitary presentation ---
 which is more convenient to work with --- of the initial model.

 Since we do not want to introduce unnecessary complexity overhead to
 readers not acquainted with double categories, we will present a
 gentle introduction to the subject. For more details we refer to
 \cite{Ehr:CS,Ehr:CS2,MM:TLRL,BMM:PTTL,BMM:SMCDC}.

 A double category contains two categorical structures, called
 \emph{horizontal} and \emph{vertical} respectively, defined over the
 same set of cells. More precisely, double categories admit the
 following na\"\i ve definition.

\begin{figure}[t]
\begin{center}
$\xymatrix{
 {o_0} \ar [r]^{h_0} \ar @{} [dr]|{A} \ar [d]_{v_0} &
 {o_1} \ar [d]^{v_1} \\
 {o_2} \ar [r]_{h_1} &
 {o_3} }$
\end{center}
\caption{Graphical representation of a cell.}
\protect\label{defcell}
\end{figure}
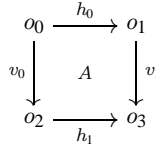

\begin{figure}
\begin{center}
$\xymatrix@+1pc{
 {\cdot} \ar[r]^{h_0} \ar[d]_{v_0} \ar@{}[rd]|{A} &
 {\cdot} \ar[r]^{h_2} \ar[d]|{v_1} \ar@{}[rd]|{B} &
 {\cdot} \ar[d]^{v_2} \\
 {\cdot} \ar[r]|{h_1} \ar[d]_{v_3} \ar@{}[rd]|{C} &
 {\cdot} \ar[r]|{h_3} \ar[d]|{v_4} \ar@{}[rd]|{D} &
 {\cdot} \ar[d]^{v_5} \\
 {\cdot} \ar[r]_{h_4} &
 {\cdot} \ar[r]_{h_5} &
 {\cdot} }$
\end{center}
\caption{Exchange law of double categories.}
\protect\label{exchlawfig}
\end{figure}
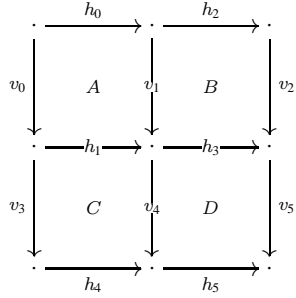

\begin{definition}[Double Category]\label{doubcat}
 A \emph{double category} $\mathcal{D}$ consists of a collection
 $o,o_0,o',\ldots$ of \emph{objects}, a collection $h,h_0,h',\ldots$
 of \emph{horizontal arrows}, a collection $v,v_0,v',\ldots$ of
 \emph{vertical arrows} and a collection $A,B,C,\ldots$ of
 \emph{double cells} (also called \emph{cells}, for short).

\begin{itemize}
\item
 Objects and horizontal arrows form the \emph{horizontal 1-category}
 $\mathcal{H}$, with identity $id_o$ for each object $o$, and
 composition $\_*\_$.

\item
 Objects and vertical arrows form also a category, called the
 \emph{vertical 1-category} $\mathcal{V}$, with identity $id_o$ for
 each object $o$, and composition $\_\cdot\_$.

\item
 Cells are assigned \emph{horizontal source} and \emph{target} (which
 are vertical arrows) and \emph{vertical source} and \emph{target}
 (which are horizontal arrows); furthermore sources and targets must
 be \emph{compatible}, in the sense that they must satisfy the
 equalities on source and target objects graphically represented by
 the square-shaped diagram in Figure~\ref{defcell}, for which we use
 the notation $A:\cell{h_0}{v_0}{v_1}{h_1}$.

\item
 Cells can be composed both horizontally ($\_*\_$) and vertically
 ($\_\cdot\_$) as follows: given $A:\cell{h_0}{v_0}{v_1}{h_1}$,
 $B:\cell{h_2}{v_1}{v_2}{h_3}$, and
 $C:\cell{h_1}{v_3}{v_4}{h_4}$, then $A *
 B:\cell{h_0*h_2}{v_0}{v_2}{h_1*h_3}$, and $A\cdot
 C:\cell{h_0}{v_0\cdot v_3}{v_1\cdot v_4}{h_4}$ are cells.
 Moreover, given a fourth cell $D:\cell{h_3}{v_4}{v_5}{h_5}$,
 horizontal and vertical compositions satisfy the following
 \emph{exchange law} (see Figure~\ref{exchlawfig}):

\[
 (A\cdot C) * (B\cdot D) = (A * B)\cdot(C * D)
\]

\noindent
 Under these rules, cells form both a horizontal category ${\cal D}^*$
 and a vertical category ${\cal D}^{\cdot}$, with identities
 $1_{v}:\cell{id_o}{v}{v}{id_{o'}}$ and
 $1^{h}:\cell{h}{id_o}{id_{o'}}{h}$, respectively, with $1^{h_0*h_1} =
 1^{h_0} * 1^{h_1}$ and $1_{v_0\cdot v_1} = 1_{v_0} \cdot 1_{v_1}$.

\item
 Furthermore, horizontal and vertical identities of identities
 coincide, \ie, $1_{id_o} = 1^{id_o}$ and the cell is simply denoted
 by $1_o$.
\end{itemize}
\end{definition}

 We shall use monoidal categories (see \eg, \cite{Mac:CWM} for
 basic definitions) for horizontal and vertical 1-categories.
 As a matter of notation,
 sequential composition and monoidal tensor product on 1-categories are denoted
 by $\_;\_$ and $\_\otimes\_$, respectively. A \emph{monoidal double
 category} is a double category together with an associative
 tensor product $\_\otimes\_$ and a unit element $e$.

 Tile logic gives a computational interpretation of (monoidal) double
 categories, where (see Figure~\ref{atilefig} for terminology):

\begin{itemize}
\item
 the objects represent the (initial/final, input/output) interfaces
 through which system components can be connected;

\item
 the arrows of $\mathcal{H}$ describe (initial/final) configurations,
 sources and targets corresponding to input and output interfaces;

\item
 the arrows of $\mathcal{V}$ are the observations (trigger/effect),
 sources and targets corresponding to initial and final interfaces;

\item
 the cells represent the possible transformations (tiles) that can be
 performed by the system.
\end{itemize}

 Thus, a cell $\cell{h_0}{v_0}{v_1}{h_1}$ says that the state
 $h_0$ can evolve to $h_1$ via an action triggered by $v_0$ and
 with effect $v_1$. The way in which observations and
 configurations of a cell are connected via their interfaces
 expresses the locality of actions, \ie, the places where triggers
 are applied to and effects are produced by.

 A basic distinction concerns whether one is interested in the
 cells or just in their borders. The second alternative has a
 more abstract flavor, in  line with behavioral equivalences, and
 corresponds to the so called \emph{flat tile logic} \cite{Bru:TL}.
 In this paper we shall concentrate on flat tiles only.

 Tile logic gives also the possibility of presenting in a constructive
 way the double category of interest. This is in some sense analogous
 to presenting a term algebra by giving only the signature: A standard
 set of rules tells how to build all the elements starting from the
 basic ones. For tile logic, the basic elements consist of: (i) the
 category $\mathcal{H}$ of configurations; (ii) the category
 $\mathcal{V}$ of observations; and (iii) the set of basic tiles (\ie,
 cells on $\mathcal{H}$ and $\mathcal{V}$). Starting from basic
 tiles, more complex tiles can be constructed by means of horizontal,
 vertical and parallel composition.  Moreover, the horizontal and
 vertical identities are always added and composed together with the
 basic tiles. All this is illustrated by the rules in
 Figure~\ref{comptilefig}, where tiles are seen as logic sequents. As
 explained in the Introduction, each operation has a precise
 computational meaning: horizontal composition coordinates the
 evolution of a context with that of its arguments; parallel
 composition models concurrent activities; vertical composition
 appends steps to computations. For both terms and tiles, the
 operation of building all the elements starting from the basic ones
 can be represented by a universal construction corresponding to a
 left adjoint.

\begin{figure}[t]
\begin{center}
 $\irule{\cell{t_0}{u}{v}{s_0}\ \ \cell{t_1}{v}{w}{s_1}}
        {\cell{t_0;t_1}{u}{w}{s_0;s_1}}$
\hfill
 $\irule{\cell{t}{u_0}{v_0}{s}\ \ \cell{s}{u_1}{v_1}{r}}
        {\cell{t}{u_0;u_1}{v_0;v_1}{r}}$
\hfill
 $\irule{\cell{t_0}{u_0}{v_0}{s_0}\ \ \cell{t_1}{u_1}{v_1}{s_1}}
        {\cell{t_0\otimes t_1}{u_0\otimes u_1}{v_0\otimes v_1}{s_0\otimes s_1}}$
\hfill $\irule{t\from o_0\to o_1\in \mathcal{H}}
        {\cell{t}{o_0}{o_1}{t}}$
\hfill $\irule{u\from o_0\to o_1\in \mathcal{V}}
        {\cell{o_0}{u}{u}{o_1}}$
\end{center}
\caption{Composition and identity rules for tile logic.}
\protect\label{comptilefig}
\end{figure}

\begin{definition}[Tile system]
 A \emph{tile system} is a tuple $\mathcal{R} =
 (\mathcal{H},\mathcal{V},N,R)$ where $\mathcal{H}$ and
 $\mathcal{V}$ are monoidal categories with the same set of
 objects $\mathbf{O}_{\mathcal{H}} = \mathbf{O}_{\mathcal{V}}$,
 $N$ is the set of rule names and $R\from N \to
 \mathbf{A}_{\mathcal{H}} \times \mathbf{A}_{\mathcal{V}} \times
 \mathbf{A}_{\mathcal{V}} \times \mathbf{A}_{\mathcal{H}}$ is a
 function such that for all $\alpha\in N$, if $R(\alpha)=\langle
 t,u,v,s\rangle$ then $t\from o_0\to o_1$, $u\from o_0\to o_2$, $v\from
 o_1\to o_3$, and $s\from o_2\to o_3$ for suitable objects $o_0$, $o_1$, $o_2$
 and $o_3$ (see Figure~\ref{gentilefig}).  We will denote such rule
 by writing $\alpha\from \cell{t}{u}{v}{s}$.
\end{definition}

\begin{figure}[t]
\begin{center}
$\xymatrix@R-1pc@C+1pc{
 {o_0} \ar[r]^{t} \ar[d]_{u} \ar@{}[rd]|{\alpha} &
 {o_1} \ar[d]^{v} \\
 {o_2} \ar[r]_{s} &
 {o_3} }$
\end{center}
\caption{A generic tile $\alpha$.}
\protect\label{gentilefig}
\end{figure}

 Depending on the chosen tile format, $\mathcal{H}$ and
 $\mathcal{V}$ can be specialized (\eg, to cartesian categories)
 and suitable \emph{auxiliary tiles} are added and composed with
 basic tiles and identities in all the possible ways. The set of
 resulting sequents (\emph{flat tiles}) define the \emph{flat
 tile logic} associated to $\mathcal{R}$. We say that
 $\cell{t}{u}{v}{s}$ is \emph{entailed} by the logic, written
 $\mathcal{R}\vdash\cell{t}{u}{v}{s}$, if the sequent
 $\cell{t}{u}{v}{s}$ can be expressed as the composition of basic
 and auxiliary tiles. Flat tiles form the cells of a suitable
 double category, which is freely generated by the tile system.

 Being interested in tile systems where configurations and
 observations are freely generated by suitable horizontal and
 vertical signatures, in what follows we shall present tile systems as
 tuples of the form $\mathcal{R}=(\Sigma,\Lambda,N,R)$. In particular,
 we shall employ categories of substitutions on $\Sigma$ and $\Lambda$
 as horizontal and vertical 1-categories. In the literature several
 tile formats have been considered
 \cite{GM:TM,FM:TFLMS,BMM:ETS,BM:CCDC,BFMM:BC}. They are all based on
 the idea of having as underlying categories of configurations and
 effects two categories that are freely generated starting from
 suitable (hyper)signatures whose operators model the basic components
 and observations, respectively.
 Varying the algebraic structure of configurations and observations,
 tiles can model many different aspects of dynamic systems,
 ranging \eg
 from synchronization of Petri net transitions \cite{BM:ZSNCCI}, to causal
 dependencies for located calculi and finitely branching approaches
 for name-passing calculi \cite{FM:TFLMS}, to actor systems
 \cite{MT:APA}, names abstraction and creation and higher order
 structures \cite{BM:CCDC}.
 A comparison between the
 various formats is out of the scope of this presentation and can be
 found in \cite{BFMM:BC}.

 Ordinary trace semantics and bisimilarity semantics can be
 extended to tiles by considering the transition system whose
 states are (possibly open) configurations and whose transitions
 are the entailed tile sequents: a tile $\cell{t}{u}{v}{s}$
 defines a transition from $t$ to $s$ with label $(u,v)$. An
 interesting question concerns suitable conditions under which
 such abstract equivalences yield congruences (\wrt\ the
 operations of the underlying horizontal structure). \emph{Tile
 decomposition} is one such condition that has a completely
 abstract formulation applicable to all tile systems.

\begin{definition}
  A tile system $\mathcal{R} = (\mathcal{H},\mathcal{V},N,R)$
  enjoys the \emph{decomposition property} if for all arrows
  $t \in \mathcal{H}$ and for all sequents
  $\cell{t}{u}{v}{s}$ entailed by~$\mathcal{R}$, then:
  (1)
  if $t = t_1;t_2$ then $\exists w\in\mathcal{V}$,
  $s_1,s_2\in\mathcal{H}$ such that $\mathcal{R}\vdash
  \cell{t_1}{u}{w}{s_1}$, $\mathcal{R}\vdash
  \cell{t_2}{w}{v}{s_2}$ and $s=s_1;s_2$;
  (2) if $t = t_1\otimes t_2$ then
  $\exists u_1,u_2,v_1,v_2\in\mathcal{V}$, $s_1,s_2\in\mathcal{H}$ such
  that $\mathcal{R}\vdash \cell{t_1}{u_1}{v_1}{s_1}$,
  $\mathcal{R}\vdash \cell{t_2}{u_2}{v_2}{s_2}$, $u=u_1\otimes
  u_2$, $v=v_1\otimes v_2$ and $s=s_1\otimes s_2$.
\end{definition}

 Condition (1) is called \emph{sequential decomposition} and condition
 (2) is called \emph{parallel decomposition}.  The decomposition property
 characterizes compositionality: It amounts to saying that if a system
 $t$ can undergo a transition $\alpha$, then for every subsystem $t_1$
 of $t$ there exists some transition $\alpha'$, such that $\alpha$ can
 be obtained by composing $\alpha'$ with a transition of the rest.

\begin{proposition}[cf.~\cite{GM:TM}]\label{decascongprop}
  If $\mathcal{R}$ enjoys the decomposition property, then tile
  bisimilarity (and also tile trace equivalence) are congruences.
\end{proposition}

 When only instantiation/contextualization are considered as
 meaningful operations of the system, then sequential decomposition is
 enough for guaranteeing the congruence of tile bisimilarity and tile
 trace equivalence \wrt\ these closure operations.

\subsection{Algebraic theories}\label{ATsec}

 An alternative presentation of the category of substitutions
 discussed in Section~\ref{Nsec} can be obtained resorting to
 \emph{algebraic theories} \cite{Law:FSAT}.

\begin{remark}
 For simplicity we illustrate here the constructions for one-sorted
 signatures. This can be extended to many sorted signatures by
 considering the free strict monoid on the set of sorts (\eg, strings
 of sorts) in place of underlined natural numbers.
\end{remark}

\begin{definition}[Algebraic theory]
 The free algebraic theory associated to a signature $\Sigma$ is the
 category $\textbf{Th}[\Sigma]$ defined below:
\begin{itemize}
\item
 its objects are `underlined' natural numbers;
\item
 the arrows from $\underline{m}$ to $\underline{n}$ are $n$-tuples of
 terms in the free $\Sigma$-algebra with (at most) $m$ canonical
 variables, and composition of arrows is term substitution. The arrows
 of $\textbf{Th}[\Sigma]$ are generated from $\Sigma$ by the inference
 rules in Figure~\ref{ThSigmafig}, modulo the axioms in
 Table~\ref{ThSigmatab}.
\end{itemize}
\end{definition}

 The category $\textbf{Th}[\Sigma]$ is isomorphic to the
 category of finite substitutions on $\Sigma$ (with canonical sets of
 variables), and the arrows from $\underline{0}$ to
 $\underline{1}$ are in bijective correspondence with the closed terms
 over $\Sigma$.

 An object $\underline{n}$ (interface) can be thought of as
 representing the $n$ (ordered) canonical variables
 $x_1,\ldots,x_n$. This allows us to denote
 $[\sbtn{t_1}{x_1},\ldots,\sbtn{t_n}{x_n}]$ just by the tuple $\langle
 t_1,\ldots,t_n\rangle$, since a standard naming of substituted
 variables can be assumed. We omit angle brackets if no confusion can
 arise.

\begin{remark}
 To avoid confusion, it must be clear that the canonical variables are
 just placeholders, \ie, their scope is only local. For example in
 $[\sbtn{f(x_1)}{x_1}]$ the two $x_1$ are different, while in
 $[\sbtn{f(x_1)}{x_1},\sbtn{g(x_1)}{x_2}]$ only the two occurrences of
 $x_1$ in $f(x_1)$ and $g(x_1)$ refer to the same placeholder. Note that
 $[\sbtn{f(x_1)}{x_1},\sbtn{g(x_2)}{x_1}]$ is inconsistent (because
 $x_1$ is assigned twice) and in fact cannot be expressed in the
 language.
\end{remark}

 The rule $\mathtt{op}$ defines basic substitutions
 $[\sbtn{f(x_1,\ldots,x_n)}{x_1}] = f(x_1,\ldots,x_n)$ for all
 $f\in\Sigma_n$.  The rule $\mathtt{id}$ yields identity substitutions
 $\langle x_1,\ldots,x_n\rangle$. The rule $\mathtt{seq}$ represents
 application of $\alpha$ to $\beta$. The rule $\mathtt{mon}$ composes
 substitutions in parallel (in $\alpha\otimes\beta$, $\alpha$ goes
 from $x_1,\ldots,x_n$ to $x_1,\ldots,x_m$, while $\beta$ goes from
 $x_{n+1},\ldots,x_{n+k}$ to $x_{m+1},\ldots,x_{m+l}$).  Three
 `auxiliary' operators (\ie, not dependent on $\Sigma$) are introduced
 that recover the cartesian structure (rules $\mathtt{sym}$,
 $\mathtt{dup}$ and $\mathtt{dis}$). The \emph{symmetry}
 $\gamma_{\underline{n},\underline{m}}$ is the permutation $\langle
 x_{n+1},\ldots,x_{n+m}, x_1,\ldots,x_n\rangle$. The \emph{duplicator}
 $\nabla_{\underline{n}} = \langle x_1,\ldots,x_n,x_1,\ldots,x_n
 \rangle$ introduces sharing and hence nonlinear substitutions.  The
 \emph{discharger} $!_{\underline{n}}$ is the empty substitution on
 $x_1,\ldots,x_n$, recovering cartesian projections.

\begin{figure}[t]
\begin{center}
$\mathtt{op}\ \
  \irule{f\in\Sigma_n}
        {f \from \underline{n} \to \underline{1}}
\hfill
  \mathtt{id}\ \
  \irule{n\in\nat}
        {id_{\underline{n}}\from \underline{n} \to \underline{n}}
\hfill
  \mathtt{seq}\ \
  \irule{\alpha \from \underline{n} \to \underline{m}\ \ \
         \beta \from \underline{m} \to \underline{k}}
        {\alpha;\beta \from \underline{n} \to \underline{k}}
\hfill
  \mathtt{mon}\ \
  \irule{\alpha \from \underline{n} \to \underline{m}\ \ \
         \beta \from \underline{k} \to \underline{l}}
        {\alpha\otimes\beta \from \underline{n+k} \to \underline{m+l}}$
\end{center}
\begin{center}
$\mathtt{sym}\ \
  \irule{n,m\in\nat}
        {\gamma_{\underline{n},\underline{m}}\from \underline{n+m} \to
\underline{m+n}}
\hfill
  \mathtt{dup}\ \
  \irule{n\in\nat}
        {\nabla_{\underline{n}}\from \underline{n} \to \underline{n+n}}
\hfill
  \mathtt{dis}\ \
  \irule{n\in\nat}
        {!_{\underline{n}}\from \underline{n} \to \underline{0}}$
\end{center}
\caption{The inference rules for the generation of $\textbf{Th}[\Sigma]$.}
\protect\label{ThSigmafig}
\end{figure}

 Let us briefly comment on the axiomatization in
 Table~\ref{ThSigmatab}. The first two rows say that
 $\textbf{Th}[\Sigma]$ is a strict monoidal category, with tensor
 product $\otimes$ and neutral element $id_{\underline{0}}$. The third
 row and the naturality axiom for symmetries (first axiom of the last
 row) say that $\textbf{Th}[\Sigma]$ is also symmetric. In particular,
 the axioms in the third row state the coherence of symmetries
 $\gamma_{\underline{n},\underline{m}}$, namely that all the
 equivalent ways of swapping the first $n$ variables with the
 following $m$ variables built out of the basic symmetry
 $\gamma_{\underline{1},\underline{1}}$ (that swaps two adjacent
 variables) are identified. The axioms in the fourth row accomplish a
 similar task for duplicators, and those in the fifth row for
 dischargers. The naturality of duplicators and dischargers (second
 and third axioms of the last row) makes $\textbf{Th}[\Sigma]$
 cartesian.

\begin{table}[t]
\begin{center}
\caption{Axiomatization of $\textbf{Th}[\Sigma]$.}
\protect\label{ThSigmatab}
\footnotesize
\begin{tabular}{llcr}
\hline
\hline
  category &
  $\alpha;(\beta;\delta) = (\alpha;\beta);\delta$ & &
  $\alpha;id_{\underline{m}} = \alpha = id_{\underline{n}};\alpha$ \\
\\
  tensor &
  $(\alpha;\alpha')\otimes(\beta;\beta') =
   (\alpha\otimes\beta);(\alpha'\otimes\beta')$ & &
  $id_{\underline{n+m}} = id_{\underline{n}}\otimes id_{\underline{m}}$ \\
  product &
  $\alpha\otimes(\beta\otimes\delta) = (\alpha\otimes\beta)\otimes\delta$ & &
  $\alpha\otimes id_{\underline{0}} = \alpha =
   id_{\underline{0}}\otimes\alpha$ \\
\\
  symmetries &
  $\gamma_{\underline{n},\underline{m+k}} =
    (\gamma_{\underline{n},\underline{m}}\otimes
    id_{\underline{k}});(id_{\underline{m}}\otimes
    \gamma_{\underline{n},\underline{k}})$ &
  $\gamma_{\underline{n},\underline{0}} = id_{\underline{n}}$ &
  $\gamma_{\underline{n},\underline{m}};
   \gamma_{\underline{m},\underline{n}} = id_{\underline{n+m}}$ \\
\\
  duplicators &
  $\nabla_{\underline{n+m}} = (\nabla_{\underline{n}}\otimes
   \nabla_{\underline{m}});(id_{\underline{n}}\otimes
   \gamma_{\underline{n},\underline{m}} \otimes id_{\underline{m}})$ & &
  $\nabla_{\underline{0}} = id_{\underline{0}}$ \\
  & $\nabla_{\underline{n}};(id_{\underline{n}}\otimes\nabla_{\underline{n}})
    = \nabla_{\underline{n}};(\nabla_{\underline{n}}\otimes
    id_{\underline{n}})$ & &
  $\nabla_{\underline{n}};\gamma_{\underline{n},\underline{n}} =
\nabla_{\underline{n}}$ \\
\\
  discharger &
  $!_{\underline{n+m}} = !_{\underline{n}} \otimes !_{\underline{m}}$ &
  $!_{\underline{0}} = id_{\underline{0}}$ &
  $\nabla_{\underline{n}};(id_{\underline{n}}\otimes
   !_{\underline{n}}) = id_{\underline{n}}$ \\
\\
  naturality &
  $(\alpha\otimes\beta);\gamma_{\underline{m},\underline{l}} =
   \gamma_{\underline{n},\underline{k}};(\beta\otimes\alpha)$ &
  $\alpha;\nabla_{\underline{m}} =
   \nabla_{\underline{n}};(\alpha\otimes\alpha)$ &
  $\alpha;!_{\underline{m}} = !_{\underline{n}}$ \\
\hline
\hline
\end{tabular}
\end{center}
\end{table}

 This presentation shows that all the auxiliary structure can be
 generated by composing together three basic constructors
 ($\gamma_{\underline{1},\underline{1}}$, $\nabla_{\underline{1}}$ and
 $!_{\underline{1}}$), \ie, it admits a finitary specification.
 Moreover, we think that this construction nicely separates the
 syntactic structure of the signature from the additional auxiliary
 structure common to all cartesian models.

 The axiomatization of $\textbf{Th}[\Sigma]$ has been exploited in
 \cite{BFMM:BC} for defining a
 taxonomy of tile formats as certain axioms or operators are omitted
 from the configuration and observation categories. In particular the
 auxiliary tiles needed in all such formats can be characterized as
 the bidimensional counterparts of symmetries, duplicators and
 dischargers. For example, let us mention that the auxiliary tiles of
 \emph{term tile logic} \cite{BMM:PTTL} (where the categories of
 configurations and observations are freely generated cartesian
 categories $\textbf{Th}[\Sigma]$ and $\textbf{Th}[\Lambda]$) are the
 commuting squares of the category $\textbf{Th}[\emptyset]$ generated
 by the empty signature.

\section{Resolution via pullbacks}\label{RVPsec}

 As we have briefly recalled in the Introduction, the construction
 of the mgu has a clear mathematical meaning: It can be
 formulated in the terminology of category theory as a well-known
 universal construction called pullback (taken in a suitable category).

 Universal constructions play a fundamental role in category
 theory, as they express the best way to accomplish a certain
 task. They usually involve a diagram that imposes certain
 constraints on the construction and then require the existence
 and uniqueness in the category of certain arrows satisfying such
 constraints, \ie, representing a possible solution to the
 problem. Among these solutions, one is of course interested in
 the optimal one (if it exists), \eg, the least upper
 bound. Categorically speaking, this is
 achieved by taking the solution that uniquely factorizes all the
 other solutions. Note that since many such optimal solutions can
 exist, any of them is completely equivalent to all the others
 (they are indeed pairwise isomorphic). Universal constructions
 allow one to recast ordinary set-theoretic constructions (\eg,
 cartesian product, disjoint sum) in a more general, abstract
 formulation that can serve as a uniform guide for catching
 analogies and pursuing comparisons between different frameworks.

\begin{definition}[Pullback]
 Given a category $\mathcal{C}$ and two arrows $h\from o_0 \to o_2$
 and $g\from o_1 \to o_2$ in $\mathcal{C}$, the \emph{pullback} of $h$ and $g$ in
 $\mathcal{C}$ is an object $o$ together with two projections
 $p_0 \from o \to o_0$ and $p_1 \from o \to o_1$ such that
 \begin{enumerate}
 \item $p_0;h = p_1;g$, and
 \item for any object $o'$ and arrows $q_0 \from o' \to o_0$ and $q_1 \from o' \to
 o_1$ such that $q_0;h = q_1;g$, then there must exist a unique arrow
 $q\from o'\to o$ such that $q;p_0 = q_0$ and $q;p_1 = q_1$.
 \end{enumerate}
\end{definition}

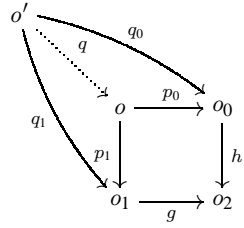
\begin{figure}[t]
\begin{center}
$\xymatrix{
 {o'} \ar@{.>}[rd]^{q} \ar@/^/[rrd]^{q_0} \ar@/_/[rdd]_{q_1} \\
 & {o} \ar[r]^{p_0} \ar[d]_{p_1} &
 {o_0} \ar[d]^{h} \\
 & {o_1} \ar[r]_{g} &
 {o_2} }$
\end{center}
\caption{The pullback of $h$ and $g$.} \protect\label{pullbackfig}
\end{figure}

 The two arrows $h$ and $g$ encode the instance of the problem,
 posing constraints on the admissible solutions. The first condition
 says that $o$, $p_0$ and $p_1$ yield a solution (called a \emph{cone}
 in category theory). The second condition states that $o$, $p_0$ and
 $p_1$ form the best solution among those contained in $\mathcal{C}$.
 The commuting diagram in Figure~\ref{pullbackfig} illustrates the
 definition (as usual in category theory, universal arrows are dotted).

\begin{example}[Pullbacks in $\Set$]
 The category $\Set$ has sets as objects and functions as arrows.
 Given $h\from X \to Z$ and $g\from Y \to Z$, then their pullback is
 the set $U = \{(x,y)\in X\times Y \mid h(x)=g(y)\}$ with the
 obvious projections on the first and second components of each
 pair in $U$.
\end{example}

 The category we are interested in is the category of substitutions on
 the signature $\Sigma$. It is well known that the mgu of a set of
 equations is an equalizer in the category of substitutions (see \eg,
 \cite{BR:CUA,Gog:WIU}, though there the authors work with the opposite
 category $\mathbf{Th}[\Sigma]^{\mathrm{op}}$ of
 $\mathbf{Th}[\Sigma]$, and therefore the mgu's are given by
 coequalizers).

\begin{definition}[Equalizer]
 Given a category $\mathcal{C}$ and two arrows $h\from o_1 \to o_2$
 and $g\from o_1 \to o_2$, the \emph{equalizer} of $h$ and $g$ in
 $\mathcal{C}$ is an object $o$ together with a projection
 $p \from o \to o_1$ such that
 \begin{enumerate}
 \item $p;h = p;g$, and
 \item for any object $o'$ and arrow $q \from o' \to o_1$
 such that $q;h = q;g$, then there must exist a unique arrow
 $q'\from o'\to o$ such that $q';p = q$.
 \end{enumerate}
\end{definition}

\begin{figure}[t]
\begin{center}
\begin{tabular}{ccc}
$\vcenter{\xymatrix{
 {o} \ar[r]^{p} &
 {o_1} \ar@/^/[r]^{h} \ar@/_/[r]_{g} &
 {o_2} \\
 {o'} \ar@{.>}[u]^{q'} \ar[ru]_{q} }}$ & &
$\vcenter{\xymatrix{
 {o} \ar@{.>}[rd]^{r} \ar@/^/[rrd]^{p} \ar@/_/[rdd]_{p} \\
 & {o'_1} \ar[r]^{p_1} \ar[d]_{p'_1} &
 {o_1} \ar[d]^{h} \\
 & {o_1} \ar[r]_{g} &
 {o_2} }}$ \\
 (a) & & (b)
\end{tabular}
\end{center}
\caption{The equalizer (a)  and the pullback of $h$ and $g$
 in the same homset (b).}
\protect\label{equalizerfig}
\end{figure}
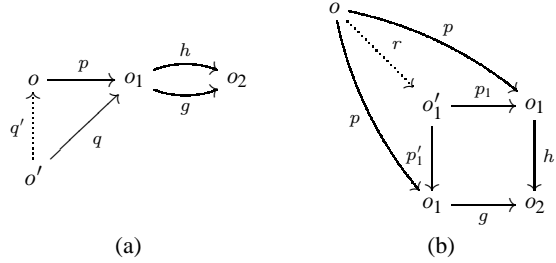

 The diagram summarizing the equalizer construction is given in
 Figure~\ref{equalizerfig}(a).

\begin{example}[Equalizers in $\Set$]
 Given $h\from X \to Z$ and $g\from X \to Z$, then their equalizer is
 the subset $U = \{x\in X \mid h(x)=g(x)\}$ of $X$ with the obvious
 inclusion $U\hookrightarrow X$ as projection.
\end{example}

 Note that, in general, for $h,g\from o_1 \to o_2$ the pullback
 of $h$ and $g$ is not
 isomorphic to their equalizer. Moreover, when both exist, the
 cone $(o,p,p)$ (obtained by taking twice the projection of the
 equalizer) uniquely factorizes through the pullback
 $(o'_1,p_1,p'_1)$ (see Figure~\ref{equalizerfig}(b)).

 For a given set of equations $\{t_1 = s_1, t_2 = s_2, \ldots, t_n=
 s_n\}$, we can consider the substitutions $\sigma =
 [\sbtn{t_1}{z_1},\sbtn{t_2}{z_2},\cdots,\sbtn{t_n}{z_n}]$ and $\sigma' =
 [\sbtn{s_1}{z_1},\sbtn{s_2}{z_2},\cdots,\sbtn{s_n}{z_n}]$, where the
 $z_i$'s are fresh variables not appearing in the $t_i$'s and
 $s_i$'s. The substitutions $\sigma$ and $\sigma'$ can be seen as
 arrows going from the set of variables appearing in the $t_i$'s and
 $s_i$'s to the set $Z = \{z_1,z_2,\ldots,z_n\}$.  To see how the
 definition of equalizer matches that of mgu, just observe that (1)
 it requires the existence of a substitution $\theta$ such that
 $\theta;\sigma = \theta;\sigma'$ and (2) the fact that $\theta$
 is the most general such substitution corresponds to the universal
 property of equalizers.

 However, in the case of logic programming, we are not really
 interested in finding the mgu of a generic set of equations, because
 we know that the variables in the head of the selected clause have
 been renamed on purpose to be different from those in the selected
 goal, \ie, they are fresh. Thus, we want to find the mgu of a set of
 equations $\{t_1 = s_1, t_2 = s_2, \ldots, t_n= s_n\}$ such that the
 variables appearing in $t_i$ and $s_j$ are disjoint for $i,j\in
 [1,n]$. Then, we can consider the substitutions $\sigma_* =
 [\sbtn{t_1}{z_1},\sbtn{t_2}{z_2},\cdots,\sbtn{t_n}{z_n}]$ and $\sigma'_*
 = [\sbtn{s_1}{z_1},\sbtn{s_2}{z_2},\cdots,\sbtn{s_n}{z_n}]$, where the
 $z_i$'s are fresh variables not appearing in the $t_i$'s and
 $s_i$'s. If we denote by $X$ the set of variables appearing in the
 $t_i$'s, by $Y$ the set of variables used in the $s_i$'s, and by $Z$
 the set $\{z_1,z_2,\ldots,z_n\}$, then we can write $\sigma_*\from X \to
 Z$ and $\sigma'_*\from Y\to Z$. Their pullback (when it exists) is
 thus given by a pair of substitutions $\psi_*\from U_*\to X$ and
 $\psi'_*\from U_*\to Y$ such that
\begin{itemize}
 \item $\psi_*;\sigma_* = \psi'_*;\sigma'_*$, and
 \item for any substitutions $\rho \from V \to X$ and $\rho' \from V
       \to Y$ such that $\rho;\sigma_* = \rho';\sigma'_*$, then there
       must exist a unique substitution $\phi \from V\to U_*$ such
       that $\phi;\psi_* = \rho$ and $\phi;\psi'_* = \rho'$.
\end{itemize}

 Since the fact that the notion of pullback of $\sigma_*$ and
 $\sigma'_*$ coincides with the notion of equalizer of $\sigma$ and
 $\sigma'$ is not completely straightforward, we illustrate below such
 a correspondence.

\paragraph{From equalizers to pullbacks.}

 Consider the arrows $\sigma\from X\cup Y \to Z$ and
 $\sigma'\from X\cup Y \to Z$ that are defined exactly as $\sigma_*$
 and $\sigma'_*$ but have different domains. Then, we know that their
 mgu is the equalizer $\theta\from U\to X\cup Y$ discussed
 above. Since $\theta$ is a substitution and since $X=\{x_1,\ldots,x_k\}$
 and $Y=\{y_1,\ldots,y_h\}$ are disjoint, then $\theta$ must have
 the form

\begin{center}
 $[\sbtn{r_1}{x_1}, \cdots, \sbtn{r_k}{x_k},
  \sbtn{r'_1}{y_1}, \cdots, \sbtn{r'_k}{y_h}]$.
\end{center}

\begin{figure}[t]
\begin{center}
\begin{tabular}{ccc}
$\vcenter{\xymatrix{
 {U} \ar[r]^{\theta} &
 {X\cup Y} \ar@/^/[r]^{\sigma} \ar@/_/[r]_{\sigma'} &
 {Z} \\
 {V} \ar@{.>}[u]^{\phi} \ar[ru]_{\rho_*} } }$ & &
$\vcenter{\xymatrix{
 & & {X} \ar[rd]^{\sigma_*} \\
 {V} \ar@/^/[rru]^{\rho} \ar@/_/[rrd]_{\rho'} \ar@{.>}[r]^{\phi} &
 {U} \ar[ru]^{\theta_X} \ar[rd]_{\theta_Y} & &
 {Z} \\
 & & {Y} \ar[ru]_{\sigma'_*} }}$
\end{tabular}
\end{center}
\caption{From equalizer to pullback.}
\protect\label{firstpartfig}
\end{figure}
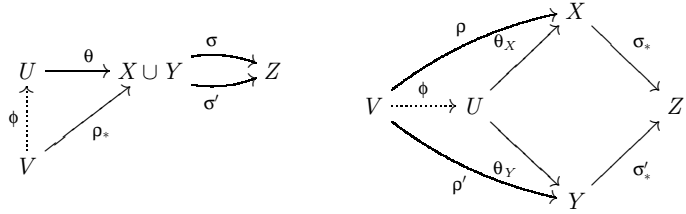

 Then, $\theta_X=[\sbtn{r_1}{x_1},\cdots,\sbtn{r_k}{x_k}]\from U \to X$
 and $\theta_Y=[\sbtn{r'_1}{y_1},\cdots,\sbtn{r'_k}{y_h}]\from U\to Y$
 satisfy $\theta_X;\sigma_* = \theta_Y;\sigma'_*$. We want to show
 that $U$, $\theta_X$ and $\theta_Y$ define a pullback of $\sigma_*$
 and $\sigma'_*$. In fact, suppose that there exist $V$ with $\rho
 \from V \to X$ and $\rho' \from V \to Y$ such that $\rho;\sigma_* =
 \rho';\sigma'_*$, then since $X$ and $Y$ are disjoint, $\rho$ and
 $\rho'$ can be combined together in a substitution $\rho_*\from V \to
 X\cup Y$ such that $\rho_*;\sigma = \rho_*;\sigma'$. By definition of
 equalizer, then there exists a unique arrow $\phi \from V \to U$ such
 that $\phi;\theta = \rho_*$. But the last condition is equivalent to
 imposing that $\phi;\theta_X = \rho$ and $\phi;\theta_Y = \rho'$
 concluding the proof. All this is illustrated in
 Figure~\ref{firstpartfig}.

\paragraph{From pullbacks to equalizers.}

\begin{figure}[t]
\begin{center}
\begin{tabular}{ccc}
$\vcenter{\xymatrix{
 & & {X} \ar[rd]^{\sigma_*} \\
 {V} \ar@/^/[rru]^{\rho_X} \ar@/_/[rrd]_{\rho_Y} \ar@{.>}[r]^{\phi_*} &
 {U'} \ar[ru]^{\psi_X} \ar[rd]_{\psi_Y} & &
 {Z} \\
 & & {Y} \ar[ru]_{\sigma'_*} }}$ & &
$\vcenter{\xymatrix{
 {U'} \ar[r]^{\psi_*} &
 {X\cup Y} \ar@/^/[r]^{\sigma} \ar@/_/[r]_{\sigma'} &
 {Z} \\
 {V} \ar@{.>}[u]^{\phi_*} \ar[ru]_{\rho_*} } }$
\end{tabular}
\end{center}
\caption{From pullback to equalizer.}
\protect\label{secondpartfig}
\end{figure}
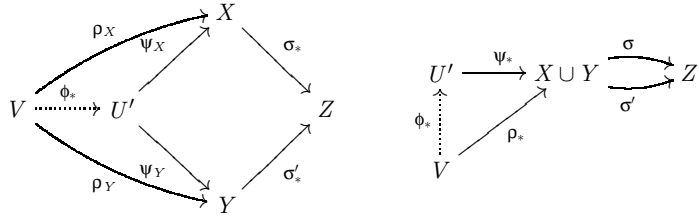

 It remains to show that to each pullback $(U',\psi_X,\psi_Y)$ of
 $\sigma_*$ and $\sigma'_*$ there corresponds an mgu (\ie, an
 equalizer) of $\sigma$ and $\sigma'$. By arguments similar to those
 employed above, it is evident that $\psi_X$ and $\psi_Y$ can be
 merged to define a substitution $\psi_*\from U'\to X\cup Y$ such that
 $\psi_*;\sigma = \psi_*;\sigma'$. Then, we must show that this
 candidate is indeed an equalizer. Thus, we assume the existence of
 $V$ and $\rho_*\from V\to X\cup Y$ such that $\rho_*;\sigma =
 \rho_*;\sigma'$. As before, we can decompose $\rho_*$ into
 $\rho_X\from V\to X$ and $\rho_Y\from V\to Y$ with $\rho_X;\sigma_* =
 \rho_Y;\sigma'_*$. By definition of pullback, then there exists a
 unique arrow $\phi_*\from V\to U'$ such that $\phi_*;\psi_X = \rho_X$
 and $\phi_*;\psi_Y = \rho_Y$, and the last two conditions are
 equivalent to the constraint $\phi_*;\psi_* = \rho_*$, concluding the
 proof.  All this is illustrated in Figure~\ref{secondpartfig}.

\paragraph{}

 Of course, it might well be the case that no such arrows $\psi$
 and $\psi'$ exist, \eg, when one tries to solve the sets $\{f(x)
 = f'(y)\}$ or $\{f(x) = x\}$ for unary operation symbols $f$ and
 $f'$.  There can also exist more solutions than one, and this is
 always the case as the names of the variables in $U$ are not
 important at all.

 The different flavors corresponding to the equalizer and the
 pullback views rely on the fact that in the equalizer
 construction we have to work with the full universe $\mathcal{X}$
 of \emph{all} variables, while in the pullback
 construction only the variables of interest for a particular mgu
 creation must be considered. Therefore, the equalizer
 approach is completely centralized, while the pullback
 construction is as much distributed as possible.
 Nevertheless, the result above proves that the two views
 are equivalent.

\section{The double category of pullbacks}\label{DCPsec}

 In this section we show that the construction of pullbacks in the
 category of substitutions can be presented in a modular way, by
 composing together a finite set of basic pullbacks.  We start by
 showing that the pullback squares in a category $\mathcal{C}$ form a
 double category. To see this, let us remind a few classical results.

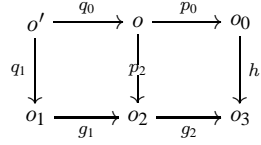
\begin{figure}
\begin{center}
$\xymatrix{
 {o'} \ar[r]^{q_0} \ar[d]_{q_1} &
 {o} \ar[r]^{p_0} \ar[d]|{p_2} &
 {o_0} \ar[d]^{h} \\
 {o_1} \ar[r]_{g_1} &
 {o_2} \ar[r]_{g_2} &
 {o_3}
}$
\end{center}
\caption{Composition of pullbacks.}
\protect\label{pbcompfig}
\end{figure}

\begin{proposition}\label{proppbcompdef}
 Given a category $\mathcal{C}$ and three arrows $h\from o_0\to o_3$,
 $g_1\from o_1 \to o_2$ and $g_2\from o_2 \to o_3$, let $(o,p_0,p_2)$
 be the pullback of $h$ and $g_2$, and let $(o',q_0,q_1)$ be the
 pullback of $p_2$ and $g_1$ (see Figure~\ref{pbcompfig}). Then,
 $(o',q_0;p_0,q_1)$ is a pullback of $h$ and $g_1;g_2$.
\end{proposition}

\begin{proposition}
 Given a category $\mathcal{C}$ and three arrows $h\from o_0\to o_3$,
 $g_1\from o_1 \to o_2$ and $g_2\from o_2 \to o_3$, let $(o,p_0,p_2)$ be the
 pullback of $h$ and $g_2$, and let $(o',q_0;p_0,q_1)$ be the pullback
 of $h$ and $g_1;g_2$. Then, $(o',q_0,q_1)$
 is a pullback of $p_2$ and $g_1$.
\end{proposition}

\begin{proposition}
 The pullback of $h\from o_0 \to o_1$ and $id_{o_1}\from o_1\to o_1$
 exists in any category. Moreover, $(o_0,id_{o_0},h)$ is a pullback
 of $h$ and $id_{o_1}$.
\end{proposition}

\begin{figure}[t]
\begin{center}
$\xymatrix{
 {o} \ar[r]^{p_0} \ar[d]_{p_1} &
 {o_0} \ar[d]^{h} \\
 {o_1} \ar[r]_{g} &
 {o_2} }$
\end{center}
\caption{Pullback square as double cell.}
\protect\label{pbsqfig}
\end{figure}
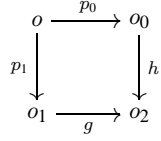

\begin{definition}[Double category of pullbacks]
 Given a category $\mathcal{C}$, the double category of pullbacks in
 $\mathcal{C}$, denoted by $\mathscr{P}(\mathcal{C})$, is defined as
 follows:
\begin{itemize}
\item
 its objects are the objects of $\mathcal{C}$;
\item
 its horizontal 1-category is $\mathcal{C}$;
\item
 its vertical 1-category is $\mathcal{C}$;
\item
 the cells are the squares $(p_0,p_1,h,g)$ (see
 Figure~\ref{pbsqfig}) such that $p_0$ and $p_1$ define a pullback
 of $h$ and $g$;
\item
 given two cells $(q,q',p',g_1)$ and $(p,p',h,g_2)$ their
 horizontal composition is the cell $(q;p,q',h,g_1;g_2)$;
\item
 given two cells $(q,q',h_1,p)$ and $(p,p',h_2,g)$ their
 vertical composition is the cell $(q,q';p',h_1;h_2,g)$.
\end{itemize}
\end{definition}

 To see that $\mathscr{P}(\mathcal{C})$ is indeed a double category,
 observe that both horizontal and vertical compositions of
 cells return pullback squares (Proposition~\ref{proppbcompdef}).
 Moreover, the trivial pullback squares $(id_{o_0},p,p,id_{o_1})$ and
 $(p,id_{o_0},id_{o_1},p)$ behave as identities \wrt\ the horizontal and
 vertical composition, respectively. The exchange law of double
 categories holds trivially, as the cells are completely identified by
 their borders.

 For the arguments presented in the previous section, it follows that
 pullbacks are a fundamental ingredient in the operational semantics,
 as they provide a characterization of the mgu construction, which
 clearly separates the goal dimension (horizontal) from the
 resolution mechanism (the vertical dimension) focusing on their
 interaction (the substitutions yielding the pullback).

 However, dealing with all this machinery at the computational level
 is too heavy, as there are infinitely many pullbacks. Therefore, a
 finitary presentation of $\mathscr{P}(\mathcal{C})$ is a main issue.

\subsection{Finitary presentation of pullbacks}\label{FPPsec}

\begin{figure}[t]
\begin{center}
$\vcenter{\xymatrix{
 {\underline{1}} \ar@{}[rd]|{R_f} &
 {\underline{n}} \ar[l]_{f} \\
 {\underline{n}} \ar[u]^{f} &
 {\underline{n}} \ar[l]^{id_{\underline{n}}} \ar[u]_{id_{\underline{n}}} }}$
\hfill
 $\vcenter{\xymatrix{
 {\underline{2}} \ar@{}[rdd]|{D_f} &
 {\underline{n+1}} \ar[l]_{f\otimes id_{\underline{1}}} \\
 & {\underline{2n}} \ar[u]_{id_{\underline{n}}\otimes f} \\
 {\underline{1}} \ar[uu]^{\nabla_{\underline{1}}} &
 {\underline{n}} \ar[l]^{f} \ar[u]_{\nabla_{\underline{n}}} }}$
\hfill $\vcenter{\xymatrix{
 {\underline{2}} \ar@{}[rrd]|{\hat{D}_f} & &
 {\underline{1}} \ar[ll]_{\nabla_{\underline{1}}} \\
 {\underline{n+1}} \ar[u]^{f\otimes id_{\underline{1}}} &
 {\underline{2n}} \ar[l]^{id_{\underline{n}} \otimes f} &
 {\underline{n}} \ar[l]^{\nabla_{\underline{n}}} \ar[u]_{f} }}$
\hfill (for all $f\in\Sigma^n$)
\end{center}
\begin{center}
$\vcenter{\xymatrix{
 {\underline{2}} \ar@{}[rd]|{R_\nabla} &
 {\underline{1}} \ar[l]_{\nabla_{\underline{1}}} \\
 {\underline{1}} \ar[u]^{\nabla_{\underline{1}}} &
 {\underline{1}} \ar[l]^{id_{\underline{1}}} \ar[u]_{id_{\underline{1}}} }} $
\hfill $\vcenter{\xymatrix{
 {\underline{2}} \ar@{}[rd]|{R_\gamma} &
 {\underline{2}} \ar[l]_{\gamma_{\underline{1},\underline{1}}} \\
 {\underline{2}} \ar[u]^{\gamma_{\underline{1},\underline{1}}} &
 {\underline{2}} \ar[l]^{id_{\underline{2}}} \ar[u]_{id_{\underline{2}}} }}$
\hfill $\vcenter{\xymatrix{
 {\underline{3}} \ar@{}[rd]|{D_\nabla} &
 {\underline{2}} \ar[l]_{\nabla_{\underline{1}}\otimes id_{\underline{1}}} \\
 {\underline{2}} \ar[u]^{id_{\underline{1}}\otimes \nabla_{\underline{1}}} &
 {\underline{1}} \ar[l]^{\nabla_{\underline{1}}}
                 \ar[u]_{\nabla_{\underline{1}}} }}$
\end{center}
\caption{The basic pullbacks.}
\protect\label{basicpbkfig}
\end{figure}
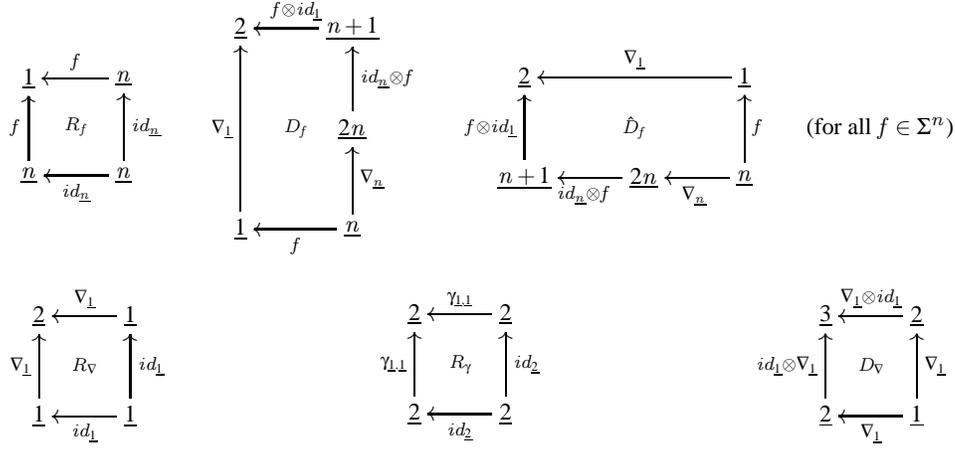

 Our first contribution consists of recovering in a finitary way the
 double category $\mathscr{P}(\textbf{Th}[\Sigma])$. We start by
 focusing on the small set of commuting squares depicted in
 Figure~\ref{basicpbkfig}. We want to show that any pullback can then
 be obtained by composing these basic squares (and horizontal and
 vertical identity pullbacks), \ie\ that the basic squares
 in Figure~\ref{basicpbkfig} form a basis for the
 generation of arbitrary pullbacks.

 There are a few points for which some explanation is worth. First of
 all, note that we have depicted the pullback cells with the direction
 of the arrows reversed with respect to the usual presentation. The
 reason for this will become much clearer in Section~\ref{TSLPsec},
 where we will show that this representation matches the intuitive
 direction of computation flow (from up to down) and also of
 internal contextualizations, which now compose to the right of the current state.
 From the point of view of the notation this is not problematic in tile logic,
 as we can assume to work with opposite categories of configuration
 and observations. As a matter of notation, the tiles in
 Figure~\ref{basicpbkfig} can be written as the sequents
 \begin{itemize}
 \item
 $R_f\from
 \cell{f}{f}{id_{\underline{n}}}{id_{\underline{n}}}$,
 \item
 $D_f\from
 \cell{f\otimes id_{\underline{1}}}{\nabla_{\underline{1}}}
      {\nabla_{\underline{n}};(id_{\underline{n}}\otimes f)}{f}$,
 \item
 $\hat{D}_f\from
 \cell{\nabla_{\underline{1}}}{f\otimes id_{\underline{1}}}
      {f}{\nabla_{\underline{n}};(id_{\underline{n}}\otimes f)}$,
 \item
 $R_\nabla\from
 \cell{\nabla_{\underline{1}}}{\nabla_{\underline{1}}}
      {id_{\underline{1}}}{id_{\underline{1}}}$,
 \item
 $R_\gamma\from
 \cell{\gamma_{\underline{1},\underline{1}}}
      {\gamma_{\underline{1},\underline{1}}}
      {id_{\underline{2}}}{id_{\underline{2}}}$, and
 \item
 $D_\nabla\from
 \cell{\nabla_{\underline{1}} \otimes id_{\underline{1}}}
      {id_{\underline{1}} \otimes \nabla_{\underline{1}}}
      {\nabla_{\underline{1}}}{\nabla_{\underline{1}}}$.
 \end{itemize}

 The second point to notice is that the cells $R_\nabla$, $R_\gamma$, and
 $D_\nabla$ do not depend on $\Sigma$, \ie, they form in some sense
 the intrinsic auxiliary structure of the pullback construction.

\begin{definition}
 We let $\mathcal{B} = \{R_\nabla, R_\gamma, D_\nabla\}$ be
 the \emph{signature-independent pullback basis}, and let
 $\mathcal{B}(f) = \{R_f ,D_f, \hat{D}_f\}$ be the \emph{pullback
 basis for the operator $f$}. Given a signature $\Sigma$,
 we call $\mathcal{B}(\Sigma) = \mathcal{B} \cup
 \bigcup_{f\in\Sigma} \mathcal{B}(f)$ the \emph{pullback
 basis for $\Sigma$}. We say that the cell
 $\cell{s}{u}{v}{t}$ is \emph{generated} by
 $\mathcal{B}(\Sigma)$, if it can be expressed as the
 parallel and sequential composition of cells in
 $\mathcal{B}(\Sigma)$ and identity cells.
\end{definition}

\begin{figure}[t]
\begin{center}
$\vcenter{\xymatrix{
 {\underline{0}} \ar@{}[rd]|{R_!} &
 {\underline{1}} \ar[l]_{!_{\underline{1}}} \\
 {\underline{1}} \ar[u]^{!_{\underline{1}}} &
 {\underline{1}} \ar[l]^{id_{\underline{1}}} \ar[u]_{id_{\underline{1}}} }}$
\hfill $\vcenter{\xymatrix{
 {\underline{3}} \ar@{}[rd]|{\hat{D}_\nabla} &
 {\underline{2}} \ar[l]_{id_{\underline{1}} \otimes \nabla_{\underline{1}}} \\
 {\underline{2}} \ar[u]^{\nabla_{\underline{1}} \otimes id_{\underline{1}}} &
 {\underline{1}} \ar[l]^{\nabla_{\underline{1}}} \ar[u]_{\nabla_{\underline{1}}} }}$
\hfill $\vcenter{\xymatrix{
 {\underline{2}} \ar@{}[rrd]|{\hat{D}'_f} & &
 {\underline{1}} \ar[ll]_{\nabla_{\underline{1}}} \\
 {\underline{1+n}} \ar[u]^{id_{\underline{1}}\otimes f} &
 {\underline{2n}} \ar[l]^{f \otimes id_{\underline{n}}} &
 {\underline{n}} \ar[l]^{\nabla_{\underline{n}}} \ar[u]_{f} }}$
\end{center}
\caption{\textbf{Q}: Are these three basic cells missing?
\textbf{A}:No.}
\protect\label{misstilesfig}
\end{figure}
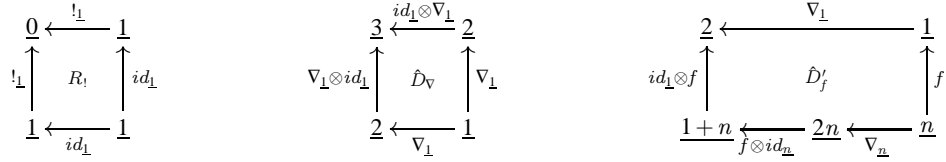

 There are some cells that one might expect to see
 in Figure~\ref{basicpbkfig} but are instead missing. Trying to
 guess the intuition of the reader, we have listed some of them in
 Figure~\ref{misstilesfig}.

 The first cell we consider is $R_!$. Its absence might be surprising,
 because there are analogous cells for all the other basic
 constructors (operators $f\in\Sigma$, symmetries and duplicators).
 But $R_!$ is not a pullback. In fact the pullback of
 $!_{\underline{1}}$ and $!_{\underline{1}}$ is
 $(\underline{2},id_{\underline{1}} \otimes !_{\underline{1}},
 !_{\underline{1}}\otimes id_{\underline{1}})$, yielding a cell that
 can in fact be obtained by composing in parallel the horizontal and
 vertical identities of $!_{\underline{1}}$ (\ie, by putting
 $\cell{id_{\underline{0}}}
       {!_{\underline{1}}}
       {!_{\underline{1}}}
       {id_{\underline{1}}}$ in parallel with
 $\cell{!_{\underline{1}}}
       {id_{\underline{0}}}
       {id_{\underline{1}}}
       {!_{\underline{1}}}$).

\begin{figure}[t]
\begin{center}
$\xymatrix@C=+2cm{
 {\underline{3}} \ar@{}[rd]|{A_1} &
 {\underline{3}} \ar[l]_{(\gamma_{\underline{1},\underline{1}}\otimes id_{\underline{1}});
                         \gamma_{\underline{2},\underline{1}}}
                 \ar@{}[rd]|{A_2} &
 {\underline{3}} \ar[l]_{\gamma_{\underline{1},\underline{1}}\otimes id_{\underline{1}}}
                 \ar@{}[rrd]|{A_3} &
 {\underline{3}} \ar[l]_{\gamma_{\underline{1},\underline{2}}} &
 {\underline{2}} \ar[l]_{id_{\underline{1}}\otimes \nabla_{\underline{1}}} \\
 {\underline{3}} \ar[u]^{(id_{\underline{1}}\otimes\gamma_{\underline{1},\underline{1}});
                         \gamma_{\underline{1},\underline{2}}}
                 \ar@{}[rd]|{A_4} &
 {\underline{3}} \ar[l]|{id_{\underline{3}}} \ar[u]|{id_{\underline{3}}}
                 \ar@{}[rrdd]|{A_5} &
 {\underline{3}} \ar[l]|{\gamma_{\underline{1},\underline{1}}\otimes id_{\underline{1}}}
                 \ar[u]|{id_{\underline{3}}} &
 {\underline{2}} \ar[l]|{\nabla_{\underline{1}}\otimes id_{\underline{1}}}
                 \ar@{}[rd]|{A_6} &
 {\underline{2}} \ar[l]|{\gamma_{\underline{1},\underline{1}}} \ar[u]_{id_{\underline{2}}} \\
 {\underline{3}} \ar[u]^{id_{\underline{1}}\otimes \gamma_{\underline{1},\underline{1}}}
                 \ar@{}[rdd]|{A_7} &
 {\underline{3}} \ar[u]|{id_{\underline{1}}\otimes \gamma_{\underline{1},\underline{1}}}
                 \ar[l]|{id_{\underline{3}}} & &
 {\underline{2}} \ar[u]|{\gamma_{\underline{1},\underline{1}}}
                 \ar@{}[rd]|{A_8} &
 {\underline{2}} \ar[u]_{id_{\underline{2}}} \ar[l]|{id_{\underline{2}}} \\
 {\underline{3}} \ar[u]^{\gamma_{\underline{2},\underline{1}}} &
 {\underline{2}} \ar[u]|{id_{\underline{1}}\otimes \nabla_{\underline{1}}}
                 \ar@{}[rd]|{A_9} &
 {\underline{2}} \ar[l]|{\gamma_{\underline{1},\underline{1}}}
                 \ar@{}[rd]|{A_{10}} &
 {\underline{1}} \ar[l]|{\nabla_{\underline{1}}} \ar[u]|{\nabla_{\underline{1}}}
                 \ar@{}[rd]|{A_{11}} &
 {\underline{1}} \ar[l]|{id_{\underline{1}}} \ar[u]_{\nabla_{\underline{1}}} \\
 {\underline{2}} \ar[u]^{\nabla_{\underline{1}} \otimes id_{\underline{1}}} &
 {\underline{2}} \ar[u]|{\gamma_{\underline{1},\underline{1}}} \ar[l]^{id_{\underline{2}}} &
 {\underline{2}} \ar[u]|{id_{\underline{2}}} \ar[l]^{id_{\underline{2}}} &
 {\underline{1}} \ar[u]|{id_{\underline{1}}} \ar[l]^{\nabla_{\underline{1}}} &
 {\underline{1}} \ar[u]_{id_{\underline{1}}} \ar[l]^{id_{\underline{1}}}
}$
\end{center}
\caption{How to compose the cell $\hat{D}_\nabla$.}
\protect\label{excompfig}
\end{figure}

\begin{figure}[t]
\begin{center}
$\xymatrix@+1pc{
 {\ } \ar@{}[rd]|{A_{11}\otimes R_{\gamma}} &
 {\ } \ar[l] \ar@{}[rd]|{A_2} &
 {\ } \ar[l] \ar@{}[rd]|{A'_2} &
 {\ } \ar[l] \\
 {\ } \ar[u] \ar@{}[rd]|{A'_4} &
 {\ } \ar[u] \ar[l] \ar@{}[rd]|{R_{\gamma}\otimes A_{11}} &
 {\ } \ar[u] \ar[l] \ar@{}[rd]|{A'_2} &
 {\ } \ar[u] \ar[l] \\
 {\ } \ar[u] \ar@{}[rd]|{A_4} &
 {\ } \ar[u] \ar[l] \ar@{}[rd]|{A_4} &
 {\ } \ar[u] \ar[l] \ar@{}[rd]|{A_{11}\otimes R_{\gamma}} &
 {\ } \ar[u] \ar[l] \\
 {\ } \ar[u] &
 {\ } \ar[u] \ar[l] &
 {\ } \ar[u] \ar[l] &
 {\ } \ar[u] \ar[l]
}$
\end{center}
\caption{How to obtain the cell $A_1$ in Figure~\ref{excompfig}.}
\protect\label{exauxfig}
\end{figure}
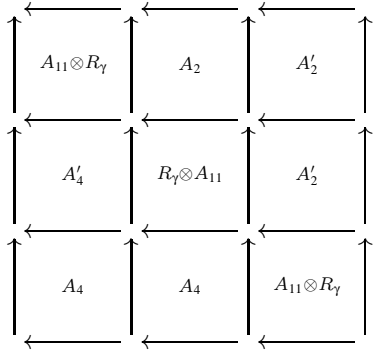

 The second cell $\hat{D}_\nabla$ defines a pullback, but it can be
 obtained by composition of other basic cells and identities, as
 illustrated in Figure~\ref{excompfig}. Let us comment on the
 composition. At the centre of the figure we find the cell $A_5 =
 D_{\nabla}$, in fact we recall that by the coherence axioms for
 duplicators (cf. Table~\ref{ThSigmatab}) we have
 $\nabla_{\underline{1}};\gamma_{\underline{1},\underline{1}} =
 \nabla_{\underline{1}}$, and by functoriality of tensor product we
 have \eg,
\[
 (id_{\underline{1}} \otimes \nabla_{\underline{1}}) ;
 (id_{\underline{1}} \otimes \gamma_{\underline{1},\underline{1}}) =
 (id_{\underline{1}} ; id_{\underline{1}}) \otimes
 (\nabla_{\underline{1}};\gamma_{\underline{1},\underline{1}}) =
 id_{\underline{1}} \otimes \nabla_{\underline{1}}.
\]
\noindent
 On the bottom-right part of the figure, we find the cells $A_8$
 and $A_{10}$ that are the horizontal and vertical identities of
 $\nabla_{\underline{1}}$, while $A_{11}$ is the trivial identity
 for the object $\underline{1}$. Then, note that $A_6 = A_9 =
 R_{\gamma}$. The cell $A_7$ is a horizontal identity, in fact by
 naturality of the symmetries, we have
\[
 \gamma_{\underline{1},\underline{1}} ;
 (id_{\underline{1}} \otimes \nabla_{\underline{1}}) =
 (\nabla_{\underline{1}} \otimes id_{\underline{1}}) ;
 \gamma_{\underline{2},\underline{1}}.
\]
\noindent
 Likewise, the cell $A_3$ is a vertical identity. Also $A_2$ and
 $A_4$ are obvious identities. The tile $A_1$ deserves more
 attention. The first thing to note is that by naturality of
 symmetries we have that
\[
 (\gamma_{\underline{1},\underline{1}}\otimes id_{\underline{1}});
 \gamma_{\underline{2},\underline{1}} =
 \gamma_{\underline{2},\underline{1}};
 (id_{\underline{1}}\otimes\gamma_{\underline{1},\underline{1}})
\]
\noindent
 and then, by coherence of symmetries, we have
\begin{eqnarray*}
 \gamma_{\underline{2},\underline{1}} & = &
 (id_{\underline{1}}\otimes\gamma_{\underline{1},\underline{1}});
 (\gamma_{\underline{1},\underline{1}}\otimes id_{\underline{1}})
 \\
 \gamma_{\underline{1},\underline{2}} & = &
 (\gamma_{\underline{1},\underline{1}}\otimes id_{\underline{1}});
 (id_{\underline{1}}\otimes\gamma_{\underline{1},\underline{1}})
\end{eqnarray*}
\noindent
 and therefore it follows that
\[
 \gamma_{\underline{2},\underline{1}};
 (id_{\underline{1}}\otimes\gamma_{\underline{1},\underline{1}}) =
 (id_{\underline{1}}\otimes\gamma_{\underline{1},\underline{1}});
 (\gamma_{\underline{1},\underline{1}}\otimes id_{\underline{1}});
 (id_{\underline{1}}\otimes\gamma_{\underline{1},\underline{1}}) =
 (id_{\underline{1}}\otimes\gamma_{\underline{1},\underline{1}});
 \gamma_{\underline{1},\underline{2}}.
\]
\noindent
 We can thus construct $A_1$ as illustrated in
 Figure~\ref{exauxfig}, where $A'_2$ is the vertical identity of
 $id_{\underline{1}}\otimes\gamma_{\underline{1},\underline{1}}$
 and $A'_4$ is the horizontal identity of
 $\gamma_{\underline{1},\underline{1}}\otimes id_{\underline{1}}$
 (for simplicity we omit to specify the borders of the
 cells, as they should be evident from the discussion above).
 To conclude that the composition in Figure~\ref{excompfig} yields
 $\hat{D}_\nabla$ we have to check that their borders are equal, and
 in fact observe that
\begin{eqnarray*}
 \gamma_{\underline{1},\underline{2}};
 (\gamma_{\underline{1},\underline{1}} \otimes id_{\underline{1}});
 (\gamma_{\underline{1},\underline{1}} \otimes id_{\underline{1}});
 \gamma_{\underline{2},\underline{1}} & = &
 \gamma_{\underline{1},\underline{2}};
 \gamma_{\underline{2},\underline{1}}  =
 id_{\underline{3}}
 \\
 \gamma_{\underline{2},\underline{1}};
 (id_{\underline{1}} \otimes \gamma_{\underline{1},\underline{1}});
 (id_{\underline{1}} \otimes \gamma_{\underline{1},\underline{1}});
 \gamma_{\underline{1},\underline{2}} & = &
 \gamma_{\underline{2},\underline{1}};
 \gamma_{\underline{1},\underline{2}}  =
 id_{\underline{3}}.
\end{eqnarray*}

 The third cell $\hat{D}'_f$ illustrated in
 Figure~\ref{misstilesfig} is a pullback, but it can be composed
 starting from $\hat{D}_f$ as shown in Figure~\ref{exffig}, where
 unnamed cells are obvious (horizontal or vertical) identities. In
 writing the border of $\hat{D}_f$ we have exploited the coherence
 axiom
\[
 \nabla_{\underline{n}};\gamma_{\underline{n},\underline{n}} =
 \nabla_{\underline{n}}.
\]
\noindent
 The cells $B_3$ and $B_2$ are identities that exploit the
 naturality of symmetries. Finally, the cell $B_1$ is obtained by
 a construction analogous to that of $A_1$, employing $R_\gamma$
 as a building block.

\begin{figure}[t]
\begin{center}
$\xymatrix@C=+2cm{
 {\underline{2}} \ar@{}[rd]|{R_\gamma} &
 {\underline{2}} \ar[l]_{\gamma_{\underline{1},\underline{1}}} &
 {\underline{2}} \ar[l]_{\gamma_{\underline{1},\underline{1}}} & &
 {\underline{1}} \ar[ll]_{\nabla_{\underline{1}}} \\
 {\underline{2}} \ar[u]^{\gamma_{\underline{1},\underline{1}}}
                 \ar@{}[rddd]|{B_3} &
 {\underline{2}} \ar[u]|{id_{\underline{2}}} \ar[l]|{id_{\underline{2}}}
                 \ar@{}[rrrd]|{\hat{D}_f} &
 {\underline{2}} \ar[u]|{id_{\underline{2}}} \ar[l]|{\gamma_{\underline{1},\underline{1}}} & &
 {\underline{1}} \ar[u]_{id_{\underline{1}}} \ar[ll]|{\nabla_{\underline{1}}} \\
 {\underline{2}} \ar[u]^{\gamma_{\underline{1},\underline{1}}} &
 {\underline{n+1}} \ar[u]|{f\otimes id_{\underline{1}}}
                   \ar@{}[rrd]|{B_2} &
 {\underline{2n}} \ar[l]|{id_{\underline{n}}\otimes f} &
 {\underline{2n}} \ar[l]|{\gamma_{\underline{n},\underline{n}}} &
 {\underline{n}} \ar[u]_{f} \ar[l]|{\nabla_{\underline{n}}} \\
 & {\underline{n+1}} \ar[u]|{id_{\underline{n+1}}}
                     \ar@{}[rd]|{B_1} &
 {\underline{1+n}} \ar[l]|{\gamma_{\underline{1},\underline{n}}} &
 {\underline{2n}} \ar[l]|{f\otimes id_{\underline{n}}} \ar[u]|{id_{\underline{2n}}} &
 {\underline{n}} \ar[l]|{\nabla_{\underline{n}}} \ar[u]|{id_{\underline{n}}} \\
 {\underline{1+n}} \ar[uu]^{id_{\underline{1}}\otimes f} &
 {\underline{1+n}} \ar[u]|{\gamma_{\underline{1},\underline{n}}} \ar[l]^{id_{\underline{1+n}}} &
 {\underline{1+n}} \ar[u]|{id_{\underline{1+n}}} \ar[l]^{id_{\underline{1+n}}}&
 {\underline{2n}} \ar[u]|{id_{\underline{2n}}} \ar[l]^{f\otimes id_{\underline{n}}} &
 {\underline{n}} \ar[u]|{id_{\underline{n}}}\ar[l]|{\nabla_{\underline{n}}}
}$
\end{center}
\caption{How to compose the cell $\hat{D}'_f$.}
\protect\label{exffig}
\end{figure}

 We hope that the few examples above can help the reader in
 understanding the compositional mechanism of basic cells, as it will
 be especially useful in Section~\ref{TSLPsec}.

 For instance, we can state a few technical lemmata that can be proved
 by tile pastings similar to the cell compositions discussed above. As
 a shorthand, for any two cells
 $A\from\cell{t}{u}{id_{\underline{n}}}{id_{\underline{n}}}$ and
 $B\from\cell{s}{v}{id_{\underline{m}}}{id_{\underline{m}}}$ with $s,v\from
 \underline{n} \to \underline{m}$, we denote by $A\lhd B$ the
 composition $(A*1^s)\cdot B = (A\cdot 1_v)*B$.

\begin{lemma}\label{lemma1}
 Given any arrow $t\from\underline{n}\to\underline{m}\in
 \mathbf{Th}[\Sigma]$ that can be obtained without using dischargers,
 the cell $\cell{t}{t}{id_{\underline{n}}}{id_{\underline{n}}}$ can be
 generated by $\mathcal{B}(\Sigma)$.
\end{lemma}

\begin{proof}
 By hypothesis, the arrow $t$ can be expressed as the parallel and
 sequential composition of arrows in $\Sigma \cup
 \{\gamma_{\underline{1},\underline{1}}, \nabla_{\underline{1}},
 id_{\underline{1}}\}$; therefore, by functoriality of tensor product,
 $t$ can be finitely decomposed as
 $\sigma_1;\sigma_2;\cdots;\sigma_l$ where $\sigma_i =
 id_{\underline{k_i}} \otimes t_i \otimes id_{\underline{m_i}}$, with
 $t_i\in \Sigma \cup \{\gamma_{\underline{1},\underline{1}},
 \nabla_{\underline{1}}, id_{\underline{1}}\}$. Then, the cell
 $\cell{t}{t}{id_{\underline{n}}}{id_{\underline{n}}}$ is just the
 (diagonal) composition $A_1\lhd A_2 \lhd \ldots \lhd A_l$, with $A_i
 = 1_{\underline{k_i}} \otimes R_{t_i} \otimes 1_{\underline{m_i}}$.
\end{proof}

 Note that by adding the cells $R_f$ just for the operators of the
 signature, then we are able to construct the analogous cells for
 generic contexts $t$.

\begin{lemma}
 Given any arrows $t\from\underline{h}\to\underline{k}$ and
 $s\from\underline{m}\to\underline{n}$ in $\mathbf{Th}[\Sigma]$, the
 cells
 $\cell{t\otimes s}{\gamma_{\underline{n},\underline{k}}}
       {\gamma_{\underline{m},\underline{h}}}{s\otimes t}$
 and
 $\cell{\gamma_{\underline{n},\underline{k}}}{t\otimes
 s}{s\otimes t}{\gamma_{\underline{m},\underline{h}}}$
 can be generated by $\mathcal{B}(\Sigma)$.
\end{lemma}
\begin{proof}
 By Lemma \ref{lemma1}, we know that the cells
 $A=\cell{\gamma_{\underline{n},\underline{k}}}
 {\gamma_{\underline{n},\underline{k}}}
 {id_{\underline{n+k}}}
 {id_{\underline{n+k}}}$ and
 $B=\cell{\gamma_{\underline{h},\underline{m}}}
 {\gamma_{\underline{h},\underline{m}}}
 {id_{\underline{h+m}}}
 {id_{\underline{h+m}}}$
 are generated by the basis $\mathcal{B}$.
 By vertically composing $B$ with the horizontal identity
 of $\gamma_{\underline{m},\underline{h}}$, we get the cell
 $C = B\cdot 1_{\gamma_{\underline{m},\underline{h}}}
 \from \cell
 {\gamma_{\underline{h},\underline{m}}}
 {id_{\underline{m+h}}}
 {\gamma_{\underline{m},\underline{h}}}
 {id_{\underline{m+h}}}$. Then, the
 cell  $\cell{t\otimes s}{\gamma_{\underline{n},\underline{k}}}
       {\gamma_{\underline{m},\underline{h}}}{s\otimes t}$
 is obtained as the composition $A*1^{s\otimes t}*C$,
 because $\gamma_{\underline{h},\underline{m}};(s\otimes
 t);\gamma_{\underline{n},\underline{k}} = t\otimes s$.
 The cell $\cell{\gamma_{\underline{n},\underline{k}}}{t\otimes
 s}{s\otimes t}{\gamma_{\underline{m},\underline{h}}}$
 can be generated by a similar construction.
\end{proof}

 The second part of the previous lemma is an instance of a
 more general result.

\begin{lemma}\label{pbduallemma}
 If the cell $\cell{s}{u}{v}{t}$ is generated by the basis
 $\mathcal{B}(\Sigma)$, then also the cell $\cell{u}{s}{t}{v}$
 does.
\end{lemma}
\begin{proof}
 Obvious, by observing that the property holds for all cells in
 $\mathcal{B}(\Sigma)$ except $D_\nabla$, for which however
 we have shown how to generate its counterpart
 $\hat{D}_\nabla$.
\end{proof}

\begin{lemma}\label{lemmanabla}
 Given any arrow $t\from\underline{m}\to\underline{n}\in
 \mathbf{Th}[\Sigma]$ that can be obtained without using dischargers,
 the cells
 $\cell{\nabla_{\underline{n}}}{t\otimes
 t}{t}{\nabla_{\underline{m}}}$ and $\cell{t\otimes
 t}{\nabla_{\underline{n}}}{\nabla_{\underline{m}}}{t}$ can be
 obtained by composition of basic cells.
\end{lemma}

\begin{theorem}
 The basis $\mathcal{B}(\Sigma)$ generates all and only pullback squares
 of $\mathbf{Th}[\Sigma]$.
\end{theorem}

\begin{proof}
 The fact that all composed cells are pullbacks is
 straightforward, as all basic tiles are pullbacks and such a
 property is preserved by the three operations of the tile model
 (horizontal and vertical sequential compositions and parallel
 composition).

 The proof that all pullbacks can be obtained in this way is more
 subtle. We exploit the fact that, in the category
 $\mathbf{Th}[\Sigma]$, whenever the pullback of $\sigma$ and $\theta$
 exists and $\sigma$ can be decomposed as $\sigma_1;\sigma_2$, then
 also the pullback of $\sigma_2$ and $\theta$ exists (because
 $\sigma_2$ is less instantiated than $\sigma$). Since each arrow
 $\sigma$ in $\mathbf{Th}[\Sigma]$ can be finitely decomposed as
 $\sigma_1;\sigma_2;\cdots;\sigma_n$ where $\sigma_i =
 id_{\underline{k_i}} \otimes t_i \otimes id_{\underline{m_i}}$, with
 $t_i\in \Sigma \cup \{\gamma_{\underline{1},\underline{1}},
 \nabla_{\underline{1}},!_{\underline{1}}, id_{\underline{1}}\}$, then
 the pullback of $\theta$ and $\sigma$, if it exists, can be computed
 stepwise. In fact, the proof is by induction on the length $n$ of a
 fixed decomposition of $\sigma$. Thus, it reduces to prove that if
 the pullback of $\theta$ and $id_{\underline{k}} \otimes t \otimes
 id_{\underline{m}}$ (with $t\in \Sigma \cup
 \{\gamma_{\underline{1},\underline{1}},
 \nabla_{\underline{1}},!_{\underline{1}}, id_{\underline{1}}\}$)
 exists, then it is generated by $\mathcal{B}(\Sigma)$. We proceed by
 case analysis on $t$ and, for each case, by induction on the length
 of the decomposition of $\theta$, exploiting the besic cells in
 $\mathcal{B}(\Sigma)$ to cover all possible combinations.
\end{proof}

\subsection{Pullbacks as tiles}\label{PATsec}

 The finitary presentation of pullbacks can be straightforwardly used
 to build a tile system that generates the double category of pullbacks.

\begin{definition}[Tile system for pullbacks]
 Given a signature $\Sigma$, we define the tile system
 $\mathcal{R}_{\mathit{PB}(\Sigma)}$ such that its horizontal category is
 $\textbf{Th}[\Sigma]^{\mathrm{op}}$, its vertical category is
 $\textbf{Th}[\Sigma]^{\mathrm{op}}$ and the basic cells are those in
 $\mathcal{B}(\Sigma)$ (see Figure~\ref{basicpbkfig} and
 remember that horizontal and vertical
 identity tiles will be freely generated in the model).
\end{definition}

 The representation theorem can then be rephrased as below.

\begin{theorem}
 A cell $\cell{t}{u}{v}{s}$ is in $\mathscr{P}(\mathcal{C})$ if and
 only if $\mathcal{R}_{\mathit{PB}(\Sigma)} \vdash \cell{t}{u}{v}{s}$.
\end{theorem}

\section{Tile systems for logic programs}\label{TSLPsec}

 The idea is to transform a logic program into a tile system which is
 able to compute the same computed answer substitutions for each
 goal. To this aim, we will exploit the tiles presented for building
 pullbacks in the category of substitutions, which provide the
 unification mechanism for goal resolution.

\subsection{From logic programs to a logic of tiles}\label{FLPTLTsec}

 The tile system that we propose can be
 sketched as follows:

\begin{definition}[Tile system for logic programming]
 Given a pure logic program $\mathcal{P}$ on the alphabet $\Sigma$, we
 denote by $\mathcal{R}_{\mathcal{P}}$ the tile system specified by
 the following rules:
\begin{itemize}
\item
 There are two basic sorts $\mathsf{t}$ (for terms) and
 $\mathsf{p}$ (for predicates). Correspondingly, the interfaces
 are elements of $\{\mathsf{t},\mathsf{p}\}^*$ (as a matter of
 notation, we let $\epsilon$ denote the empty string of sorts,
 and denote by $\mathsf{t}^n$ the string composed by $n$
 occurrences of $\mathsf{t}$, and similarly for $\mathsf{p}$).

\item
 To each functional symbol $f$ with arity $n$ in the alphabet, we
 associate an operator $\dfun{f}{\mathsf{t}^n}{\mathsf{t}}$ in the
 signature of configurations, and to each predicate symbol $p$ (here
 $\Box$ can be viewed as a nullary predicate) with arity $k$ in the
 alphabet, we associate an operator
 $\dfun{p}{\mathsf{t}^k}{\mathsf{p}}$ in the signature of
 configurations. Then, we add the symbol $\_\wedge\_ \from
 \mathsf{p}^2 \to \mathsf{p}$ for modeling conjunction. (We will show
 that, without loss of generality, the conjunction operator can be
 more conveniently defined to be associative and with unit
 $\Box$.) The configurations are the arrows (of the op-category) of the
 free cartesian category generated by the signature of configurations.

\item
 To each functional symbol $f$ with arity $n$ in the alphabet, we
 associate an operator $\dfun{f}{\mathsf{t}^n}{\mathsf{t}}$ in the
 signature of observations (note that the symbol $f$ is thus
 overloaded, since it also appears in the horizontal
 dimension;
 however, this will not create any confusion).  Then, the
 observations are the arrows (of the op-category) of the free
 cartesian category generated by the signature of observations.

\item
 To each clause
\[
 c \mathrel{\equiv} p(t_1,\ldots,t_k)\ \mathrel{:-}\
                    q_1(\vec{s}_1),\ldots,q_m(\vec{s}_m)
\]
\noindent
 (over $n$ variables $\{x_1,\ldots,x_n\}$) in the logic program we
 associate a basic tile $T_c$ in our system whose initial
 configuration is $\dfun{p}{\mathsf{t}^k}{\mathsf{p}}$ (representing
 the predicate symbol in the head of $c$), whose final
 configuration is $\dfun{ q_1(\vec{s}_1)\wedge\ldots\wedge
 q_m(\vec{s}_m)}{\mathsf{t}^n}{\mathsf{p}}$ (representing the body of
 the clause), whose trigger is the identity
 $\dfun{id_{\mathsf{p}}}{\mathsf{p}}{\mathsf{p}}$ and whose effect is
 the tuple $\dfun{\langle
 t_1,\ldots,t_k\rangle}{\mathsf{t}^n}{\mathsf{t}^k}$ (representing the
 pattern to be matched by the arguments of predicate $p$). Note that
 the body of the clause may contain variables not appearing in the
 head (\ie, some of the $x_i$'s might not appear in the $t_i$'s) and
 consequently some discharger will be used in the effect of the
 tile. Moreover, since the same variable can be used more than once,
 duplicators can also be necessary (this is to remark the difference
 between the tupling
 $\langle t_1,\ldots,t_k\rangle$ and the tensorial product $t_1\otimes
 \cdots\otimes t_k$, as the former involves duplicators for
 expressing variable sharing).
 Since we take the op-categories the direction of all arrows is
 reversed \wrt\ the standard representation (see the tile $T_c$ in
 Figure~\ref{genclausetile}).

\item
 Finally, we add the basic tiles contained in
 $\mathcal{R}_{\mathit{PB}(\Sigma)}$ (see Figure~\ref{basicpbkfig}) for
 building pullbacks. We recall that just three of them depend on the alphabet
 under consideration, while the other three are common to all
 programs, \ie, they can be considered auxiliary to the framework.
\end{itemize}
\end{definition}

\begin{remark}
 When it is obvious from the context, we shall abuse the notation by
 avoiding to specify the involved sorts in the subscripts of $id$,
 $\nabla$, $\gamma$ and $!$, writing just the numbers of involved
 arguments (\eg, instead of $\gamma_{\mathsf{tp},\mathsf{p}}$ we shall
 write $\gamma_{\underline{2},\underline{1}}\from \mathsf{tpp} \to
 \mathsf{ptp}$).
\end{remark}

\begin{figure}[t]
\begin{center}
$\xymatrix@C=2cm{
 {\mathsf{p}} \ar@{}[rd]|{T_c} &
 {\mathsf{t}^k} \ar[l]_{p} \\
 {\mathsf{p}} \ar[u]^{id} &
 {\mathsf{t}^n} \ar[l]^{q_1(\vec{s}_1)\wedge\ldots\wedge q_m(\vec{s}_m)}
                \ar[u]_{\langle t_1,\ldots,t_k\rangle}
}$
\end{center}
\caption{}
\protect\label{genclausetile}
\end{figure}

 We can assume the operator $\wedge$ to be associative and with unit
 $\Box$ because all the basic tiles associated to the clauses have an
 identity as trigger. This, together with the fact that they are the
 only rewrite rules involving predicate symbols, means that rewrites
 are always enabled for predicates nested in conjunctions. For example
 in the expression $q_1(\vec{s}_1)\wedge\ldots\wedge q_m(\vec{s}_m)$
 it is not important the way in which the $q_i$'s are conjoined, as
 their evolutions do not interact with the `tree' of
 conjunctions. Thus, $q_0 \wedge (q_1 \wedge q_2)$ is equivalent to
 $(q_0 \wedge q_1) \wedge q_2$. Moreover,  we make the special symbol $\Box$
 be the unit for $\wedge$. These assumptions do not
 alter the `behavior' of the system, but allow us to simplify the
 notation and the presentation of main results.

 The intuition is that for each goal, we can compute a refutation in the
 tile system by starting from the associated configuration and
 constructing a tile whose final configuration is the empty
 goal (possibly in parallel with some dischargers that act
 as placeholders for the free variables in the computed answer
 substitution), \ie, the final configurations must have the
 form $\Box\otimes !_{\underline{n}}$ (without monoidality of
 $\wedge$ we should have considered as final configurations
 for termination any possible finite conjunction of empty
 goals). The effect of such a tile corresponds to the computed answer
 substitution.  The tiles with initial input interface $\mathsf{p}$
 and final configuration $\Box\otimes !_{\underline{n}}$ for $n\in
 \nat$ are called \emph{refutation tiles}.

 The following example should illustrate how the tile system can
 simulate logic programming computations.

\begin{figure}[t]
\hrule \ \\
 \centerline{Program clauses translation.}
\begin{center}
$\vcenter{\xymatrix@C=2cm{
 {\mathsf{p}} \ar@{}[rd]|{T_{c_1}} &
 {\mathsf{t}^2} \ar[l]_{p} \\
 {\mathsf{p}} \ar[u]^{id} &
 {\mathsf{t}^2} \ar[l]^{(\nabla_{\underline{1}}\otimes
                         id_{\underline{1}});
                        (q \otimes r);\wedge}
                \ar[u]_{f\otimes id_{\underline{1}}}
}}$
\hfill
$\vcenter{\xymatrix@C=2cm{
 {\mathsf{p}} \ar@{}[rd]|{T_{c_2}} &
 {\mathsf{t}^2} \ar[l]_{r} \\
 {\mathsf{p}} \ar[u]^{id} &
 {\epsilon} \ar[l]^{\Box} \ar[u]_{a\otimes a}
}}$
\hfill
$\vcenter{\xymatrix@C=2cm{
 {\mathsf{p}} \ar@{}[rd]|{T_{c_3}} &
 {\mathsf{t}^1} \ar[l]_{q} \\
 {\mathsf{p}} \ar[u]^{id} &
 {\epsilon} \ar[l]^{\Box} \ar[u]_{b}
}}$
\end{center}
\hrule \ \\ \ \\
\centerline{Auxiliary pullback tiles associated to the signature.}
\begin{center}
$\vcenter{\xymatrix@C-1pc{
 {\mathsf{t}} \ar@{}[rd]|{R_a} &
 {\epsilon} \ar[l]_{a} \\
 {\epsilon} \ar[u]^{a} &
 {\epsilon} \ar[l]^{id_{\underline{0}}} \ar[u]_{id_{\underline{0}}} }}$
\hfill
 $\vcenter{\xymatrix@C-1pc{
 {\mathsf{t}^2} \ar@{}[rd]|{D_a} &
 {\mathsf{t}} \ar[l]_{a\otimes id_{\underline{1}}} \\
 {\mathsf{t}} \ar[u]^{\nabla_{\underline{1}}} &
 {\epsilon} \ar[l]^{a} \ar[u]_{a} }}$
\hfill $\vcenter{\xymatrix@C-1pc{
 {\mathsf{t}^2} \ar@{}[rd]|{\hat{D}_a} &
 {\mathsf{t}} \ar[l]_{\nabla_{\underline{1}}} \\
 {\mathsf{t}} \ar[u]^{a\otimes id_{\underline{1}}} &
 {\epsilon} \ar[l]^{a} \ar[u]_{a} }}$
\hfill
$\vcenter{\xymatrix@C-1pc{
 {\mathsf{t}} \ar@{}[rd]|{R_b} &
 {\mathsf{t}} \ar[l]_{b} \\
 {\mathsf{t}} \ar[u]^{b} &
 {\epsilon} \ar[l]^{id_{\underline{0}}} \ar[u]_{id_{\underline{0}}} }}$
\hfill
 $\vcenter{\xymatrix@C-1pc{
 {\mathsf{t}^2} \ar@{}[rd]|{D_b} &
 {\mathsf{t}} \ar[l]_{b\otimes id_{\underline{1}}} \\
 {\mathsf{t}} \ar[u]^{\nabla_{\underline{1}}} &
 {\epsilon} \ar[l]^{b} \ar[u]_{b} }}$
\hfill
$\vcenter{\xymatrix@C-1pc{
 {\mathsf{t}^2} \ar@{}[rd]|{\hat{D}_b} &
 {\mathsf{t}} \ar[l]_{\nabla_{\underline{1}}} \\
 {\mathsf{t}} \ar[u]^{b\otimes id_{\underline{1}}} &
 {\epsilon} \ar[l]^{b} \ar[u]_{b} }}$
\end{center}
\begin{center}
$\vcenter{\xymatrix{
 {\mathsf{t}} \ar@{}[rd]|{R_f} &
 {\mathsf{t}} \ar[l]_{f} \\
 {\mathsf{t}} \ar[u]^{f} &
 {\mathsf{t}} \ar[l]^{id_{\underline{1}}} \ar[u]_{id_{\underline{1}}} }}$
\hfill
 $\vcenter{\xymatrix{
 {\mathsf{t}^2} \ar@{}[rd]|{D_f} &
 {\mathsf{t}^2} \ar[l]_{f\otimes id_{\underline{1}}} \\
 {\mathsf{t}} \ar[u]^{\nabla_{\underline{1}}} &
 {\mathsf{t}} \ar[l]^{f}
                 \ar[u]_{\nabla_{\underline{1}};
                         (id_{\underline{1}}\otimes f)} }}$
\hfill $\vcenter{\xymatrix@C+1pc{
 {\mathsf{t}^2} \ar@{}[rd]|{\hat{D}_f} &
 {\mathsf{t}} \ar[l]_{\nabla_{\underline{1}}} \\
 {\mathsf{t}^2} \ar[u]^{f\otimes id_{\underline{1}}} &
 {\mathsf{t}} \ar[l]^{\nabla_{\underline{1}};
                         (id_{\underline{1}} \otimes f)}
                 \ar[u]_{f} }}$
\end{center}
\hrule \ \\ \ \\
\centerline{Auxiliary pullback tiles independent of the
signature.}
\begin{center}
$\vcenter{\xymatrix{
 {\mathsf{t}^2} \ar@{}[rd]|{R_\nabla} &
 {\mathsf{t}} \ar[l]_{\nabla_{\underline{1}}} \\
 {\mathsf{t}} \ar[u]^{\nabla_{\underline{1}}} &
 {\mathsf{t}} \ar[l]^{id_{\underline{1}}} \ar[u]_{id_{\underline{1}}} }} $
\hfill $\vcenter{\xymatrix{
 {\mathsf{t}^2} \ar@{}[rd]|{R_\gamma} &
 {\mathsf{t}^2} \ar[l]_{\gamma_{\underline{1},\underline{1}}} \\
 {\mathsf{t}^2} \ar[u]^{\gamma_{\underline{1},\underline{1}}} &
 {\mathsf{t}^2} \ar[l]^{id_{\underline{2}}} \ar[u]_{id_{\underline{2}}} }}$
\hfill $\vcenter{\xymatrix{
 {\mathsf{t}^3} \ar@{}[rd]|{D_\nabla} &
 {\mathsf{t}^2} \ar[l]_{\nabla_{\underline{1}}\otimes id_{\underline{1}}} \\
 {\mathsf{t}^2} \ar[u]^{id_{\underline{1}}\otimes \nabla_{\underline{1}}} &
 {\mathsf{t}} \ar[l]^{\nabla_{\underline{1}}}
                 \ar[u]_{\nabla_{\underline{1}}} }}$
\end{center}
\hrule
\caption{The tile system associated to the logic program
$\mathcal{P}$.} \protect\label{protilefig}
\end{figure}
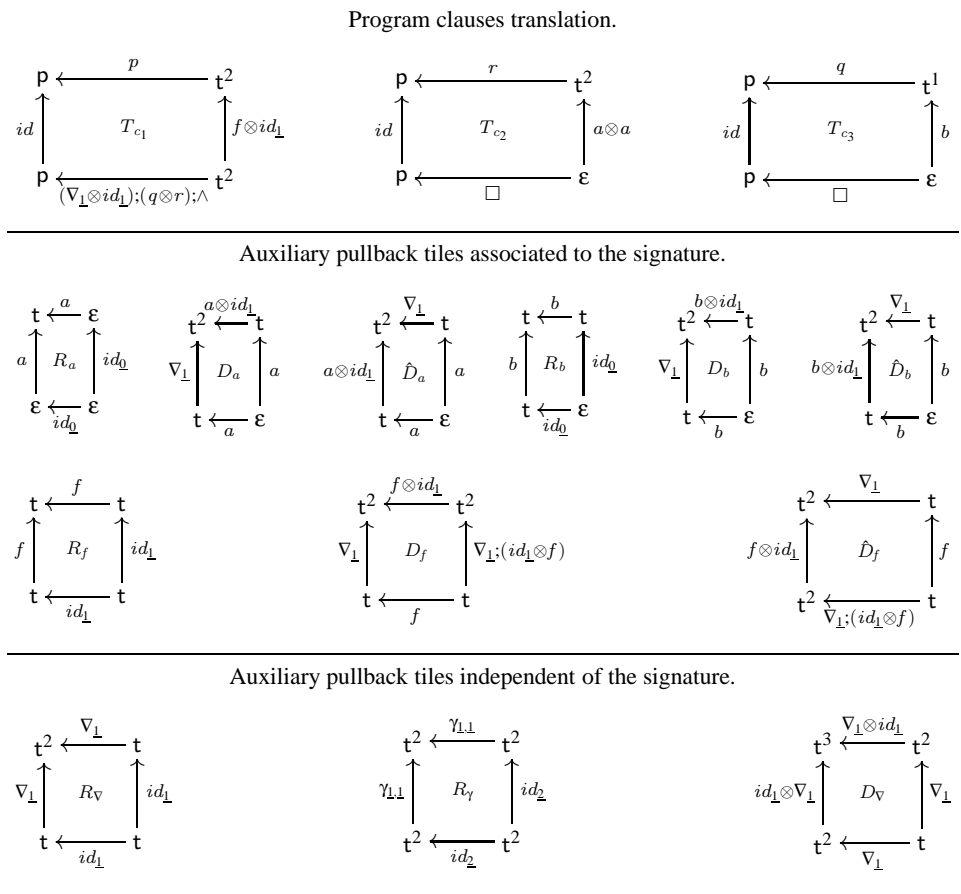

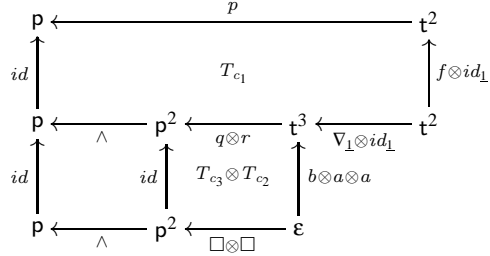
\begin{figure}[t]
\begin{center}
$\vcenter{\xymatrix@C+1pc{
 {\mathsf{p}} \ar@{}[rrrd]|{T_{c_1}} & & &
 {\mathsf{t}^2} \ar[lll]_{p} \\
 {\mathsf{p}} \ar[u]^{id} &
 {\mathsf{p}^2} \ar[l]^{\wedge} &
 {\mathsf{t}^3} \ar[l]^{q \otimes r} &
 {\mathsf{t}^2} \ar[l]^{\nabla_{\underline{1}}\otimes
                         id_{\underline{1}}}
                \ar[u]_{f\otimes id_{\underline{1}}} \\
 {\mathsf{p}} \ar[u]^{id} &
 {\mathsf{p}^2} \ar[l]^{\wedge} \ar[u]^{id}
                \ar@{}[ru]|{T_{c_3}\otimes T_{c_2}} &
 {\epsilon} \ar[l]^{\Box \otimes \Box} \ar[u]_{b\otimes a\otimes a}
}}$
\end{center}
\caption{The incomplete derivation for $p$.}
\protect\label{partcompfig}
\end{figure}

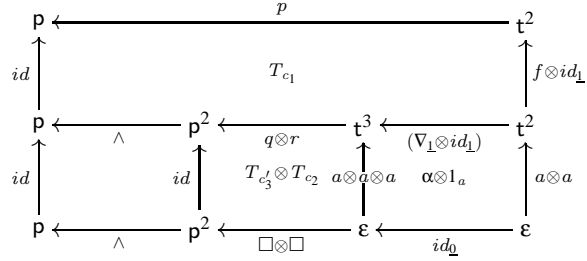
\begin{figure}[t]
\begin{center}
$\vcenter{\xymatrix@C+2pc{
 {\mathsf{p}} \ar@{}[rrrd]|{T_{c_1}} & & &
 {\mathsf{t}^2} \ar[lll]_{p} \\
 {\mathsf{p}} \ar[u]^{id} &
 {\mathsf{p}^2} \ar[l]^{\wedge} &
 {\mathsf{t}^3} \ar[l]^{q \otimes r} &
 {\mathsf{t}^2} \ar[l]^{(\nabla_{\underline{1}}\otimes
                         id_{\underline{1}})}
                \ar[u]_{f\otimes id_{\underline{1}}} \\
 {\mathsf{p}} \ar[u]^{id} &
 {\mathsf{p}^2} \ar[l]^{\wedge} \ar[u]^{id}
                \ar@{}[ru]|{T_{c'_3}\otimes T_{c_2}} &
 {\epsilon} \ar[l]^{\Box \otimes \Box} \ar[u]|{a\otimes a\otimes a}
                \ar@{}[ru]|{\alpha\otimes 1_a} &
 {\epsilon} \ar[l]^{id_{\underline{0}}} \ar[u]_{a\otimes a}
}}$
\end{center}
\caption{The refutation for $p$.}
\protect\label{compcompfig}
\end{figure}

\begin{figure}[t]
\begin{center}
$\vcenter{\xymatrix{
 {\mathsf{t}^2} \ar@{}[rd]|{\hat{D}_a} &
 {\mathsf{t}} \ar[l]_{\nabla_{\underline{1}}} \\
 {\mathsf{t}} \ar[u]^{a\otimes id_{\underline{1}}} &
 {\epsilon} \ar[l]^{a} \ar[u]_{a} \\
 {\epsilon} \ar[u]^{a}
                \ar@{}[ru]|{R_a} &
 {\epsilon} \ar[l]^{id_{\underline{0}}} \ar[u]_{id_{\underline{0}}}
}}$
\hspace{1cm}
$\vcenter{\xymatrix{
 {\mathsf{t}^2} \ar@{}[rd]|{\hat{D}'_a} &
 {\mathsf{t}} \ar[l]_{\nabla_{\underline{1}}} \\
 {\mathsf{t}} \ar[u]^{id_{\underline{1}}\otimes a} &
 {\epsilon} \ar[l]^{a} \ar[u]_{a} \\
 {\epsilon} \ar[u]^{a}
                \ar@{}[ru]|{R_a} &
 {\epsilon} \ar[l]^{id_{\underline{0}}} \ar[u]_{id_{\underline{0}}}
}}$
\end{center}
\caption{Two ways for composing the tile $\alpha$ of
 Figure~\protect\ref{compcompfig}.}
\protect\label{twowayfig}
\end{figure}
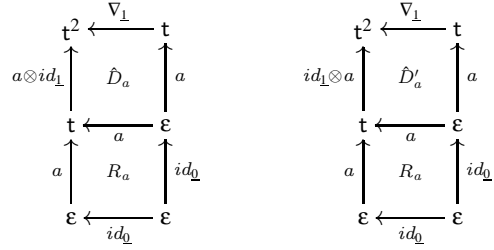

\begin{example}
 Let us consider the simple
 alphabet consisting of constants $a$ and $b$, unary function symbol
 $f$, unary predicate $q$ and binary predicates $p$ and $r$.

 Given the logic program $\mathcal{P}$ defined by the three clauses

\begin{alltt}
 \(c\sb{1} \equiv\) p(f(X1),X2) :-  q(X1), r(X1,X2).
 \(c\sb{2} \equiv\) r(a,a).
 \(c\sb{3} \equiv\) q(b).
\end{alltt}

\noindent
 the corresponding tile system is illustrated in
 Figure~\ref{protilefig}. The tiles in the first row are those
 associated to the three clauses of the program. The tiles in the
 second row are the basic pullbacks associated to the constants $a$, $b$
 of the alphabet, while the tiles in the third row are the basic
 pullbacks associated to the unary function symbol $f$ of the
 alphabet. The tiles in the fourth row are the auxiliary tiles common
 to all representations of logic programs. Note that, once the
 signature of terms is fixed, then all the auxiliary tiles
 are fixed, and the tiles for representing the logic
 program are in bijection with the clauses of the program.

 Now suppose one wants to compute the goal
 $\mathrel{?-}\ p(x_1,x_2)$. The idea is to compute all possible tiles
 that have $p\from \mathsf{t}^2 \to \mathsf{p}$ as initial
 configuration and the empty goal as conclusion. The effect
 of such tiles should in fact correspond to the computed answer
 substitutions of the program execution on the given goal. It is easy
 to argue that no such tile exists for the given goal. In fact, the
 only tile having $p$ as initial configuration is $T_{c_1}$ that leads
 to the configuration $(\nabla_{\underline{1}}\otimes
 id_{\underline{1}});(q\otimes r);\wedge$. Then $T_{c_3}$ and
 $T_{c_2}$ can be (concurrently) applied respectively to $q$ and to
 $r$, but the computation cannot be completed, as the coordination of
 the two resolutions is not possible. In fact the pullback of
 $b\otimes a$ and $\nabla_{\underline{1}}$ does not exist, and hence
 also the pullback of $b\otimes a\otimes a$ and
 $(\nabla_{\underline{1}}\otimes id_{\underline{1}})$ does not exist
 as well. The partial computation is illustrated in
 Figure~\ref{partcompfig}.

 If the third clause $c_3$ is replaced by $c'_3 \equiv \mathtt{q(a)}$,
 then we can compute the tile refutation illustrated in
 Figure~\ref{compcompfig}, where the tile $\alpha\from
 \cell{\nabla_{\underline{1}}}{a\otimes a}{a}{id_{\underline{0}}}$ can
 be obtained in any of the two ways illustrated in
 Figure~\ref{twowayfig}. The computed
 answer substitution $f(a)\otimes a$ (representing
 $[\sbtn{f(a)}{x_1},\sbtn{a}{x_2}]$) is given by the effect of the
 composed tile. Note that $c'_3$ and $c_2$ can be applied concurrently, \ie,
 the order in which they are applied is not relevant and moreover,
 they can also be performed in parallel, their outputs being
 coordinated by means of the tile $\alpha$. The two ways of building
 $\alpha$ show that the coordination mechanism does not depend on the
 order of execution of $T_{c'_3}$ and $T_{c_2}$, which is in fact
 immaterial.
\end{example}

 Notice that the use of tiles, thanks to its abstract flavour,
 completely frees the user from managing fresh variables, taking
 care in an automatic way of all the problems connected to name
 handling via the use of local placeholders.

\subsection{From clauses to tiles}\label{FCTTsec}

 We try to explain here informally the intuition that lies behind the
 definition of $\mathcal{R}_{\mathcal{P}}$. Basically, it is strictly
 related to the idea of building an \textsc{lts} out of a reduction
 system to study the interactions between composed components. From
 one point of view, it is evident that the reduction system of goal
 resolution summarized in Section~\ref{OSsec} considers the whole goal
 as an atomic entity, whose parts must all be coordinated. From this
 point of view, the clauses define the basic reduction steps that can
 be conveniently instantiated and contextualized. Indeed, the
 reduction perspective of logic programs has been investigated in
 \cite{CM:ASSTS}. However, to accomplish this view, one usually
 assumes to start with a set of variables large enough to contain all
 names that will be needed by all clause instances used in the
 refutation, as their dynamic creation cannot be modeled. This is a
 very strong assumption that somehow clashes against the desirable
 constructive presentation of computation, where fresh
 variables can be introduced by need.

 In \cite{Sew:RRBC,LM:DBCRS} it is suggested that instead of studying
 the behavior of a process in all possible contexts, the basic
 reduction rules of the system can be used to catch the least set of
 contexts that should be considered. This is obtained by considering
 all subterms of the sources of reduction rules. For example, if a
 reduction rewrites $f(g(a))$ to $h(b)$, then the essential contexts
 are $f(\_)$ and $f(g(\_))$, but not $h(\_)$, because only by
 embedding a term within these contexts a reduction may happen (unless
 it is already enabled inside the term itself). Unfortunately, this
 task is hard to accomplish in general, as the reduction semantics for
 process calculi usually impose suitable structural axioms on the
 processes. Nevertheless, the presence of sufficiently many
 \emph{relative pushouts} in the category of states is enough for
 guaranteeing that the universal constructions exist \cite{LM:DBCRS}.

 For logic programming, the problem of contextualization is reversed
 to the problem of instantiation, and we know in advance what are the
 interesting `internal' contexts, namely the pullback
 projections. This allows us to transform all clauses (seen as
 reduction rules) by moving as much internal context as possible to
 the observational part: We separate the topmost operator of the head
 of the clause (\ie, the
 predicate symbol) from its arguments (that are moved to the observational
 part, \ie, the effect of the tile) and
 then the basic pullbacks allow us to build incrementally all the
 other decompositions (in particular, we are speaking about tiles
 $R_f$, $R_\nabla$ and $R_\gamma$).

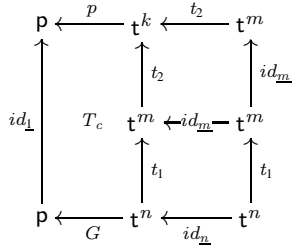
\begin{figure}[t]
\begin{center}
$\xymatrix{
 {\mathsf{p}} \ar@{}[rdd]|{T_c} &
 {\mathsf{t}^k} \ar[l]_{p} &
 {\mathsf{t}^m} \ar[l]_{t_2} \\
 & {\mathsf{t}^m} \ar[u]_{t_2} &
 {\mathsf{t}^m} \ar[l]|{id_{\underline{m}}} \ar[u]_{id_{\underline{m}}} \\
 {\mathsf{p}} \ar[uu]^{id_{\underline{1}}} &
 {\mathsf{t}^n} \ar[l]^{G} \ar[u]_{t_1} &
 {\mathsf{t}^n} \ar[l]^{id_{\underline{n}}} \ar[u]_{t_1} }$
\end{center}
\caption{Graphical proof of Proposition~\protect\ref{pt1t2prop}.}
\protect\label{pt1t2fig}
\end{figure}

\begin{proposition}\label{pt1t2prop}
 For each tile $T_c\from \cell{p}{id}{t}{G}$ and all arrows $t_1,t_2$
 such that $t = t_1;t_2$ (with $t_2$ not involving dischargers), then
 the tile $\cell{t_2;p}{id}{t_1}{G}$ is entailed by
 $\mathcal{R}_{\mathcal{P}}$.
\end{proposition}
\begin{proof}
 The proof follows from Lemma~\ref{lemma1}, \ie, from the existence of
 the tile $\cell{t_2}{t_2}{id}{id}$ that can be vertically composed
 with $\cell{id}{t_1}{t_1}{id}$ (the horizontal identity for $t_1$),
 and with $T_c$ being horizontally composed with the result (see
 Figure~\ref{pt1t2fig}).
\end{proof}

\begin{example}
 Let us consider the simple program

\begin{alltt}
 \(c\sb{1} \equiv\) sum(0,X1,X1).
 \(c\sb{2} \equiv\) sum(s(X1),X2,s(X3)) :- sum(X1,X2,X3).
\end{alltt}

\noindent
 over the signature consiting of constant $0$, unary symbol $s$ and
 ternary predicate symbol $\mathit{sum}$. The possible interactive
 decompositions of the heads of the two clauses are:
\begin{itemize}
\item
 for $c_1$: (1) $\mathit{sum}(0,x_1,x_2)$ with observation
 $\nabla_{\underline{1}}$, (2) $\mathit{sum}(x_1,x_2,x_2)$ with
 observation $0\otimes id_{\underline{1}}$, (3)
 $\mathit{sum}(x_1,x_2,x_3)$ with observation
 $0\otimes\nabla_{\underline{1}}$, and (4) $\mathit{sum}(0,x_1,x_1)$
 with observation $id_{\underline{1}}$;

\item
 for $c_2$: (1) $\mathit{sum}(s(x_1),x_2,x_3)$ with observation
 $id_{\underline{2}}\otimes s$, (2) $\mathit{sum}(x_1,x_2,s(x_3))$
 with observation $s\otimes id_{\underline{2}}$, (3)
 $\mathit{sum}(x_1,x_2,x_3)$ with observation $s\otimes
 id_{\underline{1}} \otimes s$, and finally (4)
 $\mathit{sum}(s(x_1),x_2,s(x_3))$ with observation
 $id_{\underline{3}}$.
\end{itemize}

\noindent
 Although the basic tiles of the tile system associated to the program
 just consider decompositions of kind (3), which are the most
 general, by parallel and sequential composition with (basic)
 pullback tiles, the tile logic associated to the tile system will
 entail all the other decompositions.
\end{example}

 Note that tiles allow one to move contexts along states and
 observations in a very natural and uniform way.
 The interactivity of the tile representation relies on the fact that
 the effects of basic tiles associated to the clauses must be accepted
 by the current instantiation of the matched predicate in the goal,
 otherwise the step cannot take place.

\begin{theorem}[Correspondence between (SLD-)derivations and tiles]
\label{reptheo}
 Let $\mathcal{P}$ be a logic program and $G$ a goal. Then,
\begin{enumerate}
\item
 if $\mathcal{P}\vvdash G \Rightarrow_{\sigma} G'$, then
 $\mathcal{R}_{\mathcal{P}}\vdash \cell{G}{id}{\theta}{G'}$ with
 $\theta=\sigma_{|\mathit{Var}(G)}$;

\item
 if $\mathcal{R}_{\mathcal{P}}\vdash \cell{G}{id}{\theta}{G'}$, then
 there exists $\sigma$ with $\theta=\sigma_{|\mathit{Var}(G)}$ such
 that $\mathcal{P}\vvdash G \Rightarrow^*_{\sigma} G'$.
\end{enumerate}
\end{theorem}

\begin{proof}
 The proof of point 1 proceeds by rule induction. For the `empty goal'
 rules we rely on the fact that $\Box$ is the unit for $\wedge$
 and that the vertical identities always exist. For the `atomic goal'
 we rely on the results of Section~\ref{RVPsec} on the correspondence
 between mgu's and pullbacks while applying the tile $T_{H
 \mathrel{:-} F}$ to the goal $A$. For the `conjunctive goal' rules,
 the difficulty is that $G$ and $G'$ might share some variables. In
 fact, by inductive hypothesis we can assume that
 $\mathcal{R}_{\mathcal{P}}\vdash \cell{G}{id}{\sigma}{F}$ and
 therefore we must employ the pullback tiles for propagating $\sigma$
 to $G'$. This can be done by exploiting the tiles $\hat{D}_f$ and
 $D_{\nabla}$.

 For proving the point 2, we fix a decomposition of
 $\cell{G}{id}{\theta}{G'}$ in terms of basic tiles of
 $\mathcal{R}_{\mathcal{P}}$ and then we proceed by induction on the
 number of tiles $T_c$ used for building $\cell{G}{id}{\theta}{G'}$.
\end{proof}

 Note that a tile can represent in general a whole sequence of derivation
 steps.

\begin{corollary}
  Let $\mathcal{P}$ be a logic program and $G$ a goal. Then,
\begin{enumerate}
\item
 if $\mathcal{P}\vdash_\sigma G$, then
 $\mathcal{R}_{\mathcal{P}}\vdash \cell{G}{id}{\theta}{\Box \otimes
 !_{\underline{n}}}$ with $\theta=\sigma_{|\mathit{Var}(G)}$ and $n$
 the number of free variables in $\theta$;

\item
 if $\mathcal{R}_{\mathcal{P}}\vdash \cell{G}{id}{\theta}{\Box \otimes
 !_{\underline{n}}}$, then
 there exists $\sigma$ with
 $\theta=\sigma_{|\mathit{Var}(G)}$ such
 that $\mathcal{P}\vdash_\sigma G$.
\end{enumerate}
\end{corollary}

\subsection{Recovering ordinary semantics}\label{ROSsec}

 From the tile system $\mathcal{R}_{\mathcal{P}}$ we are able to recover
 several well-known semantics for logic programs.

\begin{itemize}
\item
 The \emph{least Herbrand model}, which gives the ordinary
 model-theoretic semantics for logic programs \cite{EK:SPLPL}, is
 given by refutation tiles whose initial configuration is a ground
 atomic goal:

\[
 \mathrm{Op}_1(\mathcal{P}) =
 \{ A\from \epsilon \to \mathsf{p} \mid
 \mathcal{R}_{\mathcal{P}}\vdash
 \cell{A}
      {id_{\underline{1}}}
      {!_{\underline{n}}}
      {\Box\otimes !_{\underline{n}}} \}.
\]

\item
 The \emph{correct answer substitutions} are given by the instances of
 initial configurations of refutation tiles that are (possibly
 non-ground) atomic goals:

\[
 \mathrm{Op}_2(\mathcal{P}) =
 \{ A\from \mathsf{t}^k \to \mathsf{p} \mid
 \mathcal{R}_{\mathcal{P}}\vdash
 \cell{A}
      {id_{\underline{1}}}
      {id_{\underline{k}} \otimes !_{\underline{k+n}}}
      {\Box\otimes !_{\underline{n}}} \}.
\]

\item
 The \emph{computed answer substitutions}, which define a useful
 semantic framework for addressing compositionality, concrete
 observables and program analysis \cite{FLMP:DMOBLL,BGLM:SSA}, can be
 immediately obtained by considering the refutation tiles with a
 single predicate as initial configuration:

\[
 \mathrm{Op}_3(\mathcal{P}) =
 \{ \theta;p\from \mathsf{t}^k \to \mathsf{p} \mid
 p\in\Sigma_{\Pi},\
 \mathcal{R}_{\mathcal{P}}\vdash
 \cell{p}
      {id_{\underline{1}}}
      {\theta\otimes !_{\underline{n}}}
      {\Box\otimes !_{\underline{k+n}}} \}.
\]

\item
 The \emph{resolvents} can be obtained by considering also non
 refutation tiles:

\[
 \mathrm{Op}_4(\mathcal{P}) =
 \{ (\theta;p,G) \mid
 p\in\Sigma_{\Pi},\
 \mathcal{R}_{\mathcal{P}}\vdash
 \cell{p}
      {id_{\underline{1}}}
      {\theta}
      {G} \}.
\]
\end{itemize}

 All the correspondences above follow as easy corollaries to the
 representation Theorem~\ref{reptheo}.

\subsection{Goal compositionality}\label{GCsec}

 Though compositionality issues for the classical semantics have been
 extensively studied in the literature, we want to focus here on
 compositionality of goals \wrt\ the two main operations discussed in
 the Introduction, namely instantiation and conjunction
 (AND-compositionality). We focus on goal equivalence for a given
 program $\mathcal{P}$; thus,
 the main questions are: (1) When are two goals equivalent? (2) Is
 equivalence a congruence?

 Inspired by the connections with the area of process description
 calculi that motivated our approach, and having at hand an
 established theory developed for trace equivalence and bisimilarity
 in the tile setting, the natural step is to try to apply general
 existing techniques to our special case.  Therefore we can answer
 question (1) by defining the two equivalences:

\begin{enumerate}
\renewcommand{\theenumi}{\alph{enumi}}
\item
 $G \simeq_{\mathcal{P}} G'$ if $\mathcal{T}_{\mathcal{P}}(G) =
 \mathcal{T}_{\mathcal{P}}(G')$, where $\mathcal{T}_{\mathcal{P}}(G)
 \stackrel{\mathrm{def}}{=}\{\theta \mid
 \mathcal{R}_{\mathcal{P}}\vdash \cell{G}{id_{\underline{1}}}
 {\theta}{\Box\otimes id_{\underline{n}}} \}$.

\item
  $G \cong_{\mathcal{P}} G'$ if $G$ and $G'$ are tile bisimilar in
  $\mathcal{R}_{\mathcal{P}}$.
\end{enumerate}

 These equivalences are reminiscent of the analogous notions on the
 processes of a fixed process description calculus modeled with
 tiles. The interactive part of the underlying tile system tells what
 can be observed during the computation, and then the equivalences
 arise naturally as behavior-based concepts.

 Now, question (2) corresponds to ask whether $G \simeq G'$ implies
 that $G\wedge F \simeq G'\wedge F$ and $\sigma;G \simeq \sigma;G'$
 for all $F$ and $\sigma$ or not (the same for $\cong$).

 Though proving directly these properties is not too complicate, we can exploit
 Proposition~\ref{decascongprop} and just prove that the tile system
 $\mathcal{R}_{\mathcal{P}}$ enjoys the decomposition property for any
 logic program $\mathcal{P}$.

\begin{proposition}
 For any logic program $\mathcal{P}$, the corresponding tile system
 $\mathcal{R}_{\mathcal{P}}$ enjoys the sequential decomposition
 property.
\end{proposition}

\begin{proof}
 We want to prove that for any goal $\sigma;G$ and tile
 $\mathcal{R}_{\mathcal{P}} \vdash
 \cell{\sigma;G}{id_{\underline{1}}}{\theta}{F}$, there exist
 $\theta'$, $\sigma'$ and $F'$ such that $\mathcal{R}_{\mathcal{P}}
 \vdash \cell{G}{id_{\underline{1}}}{\theta'}{F'}$ and
 $\mathcal{R}_{\mathcal{P}} \vdash
 \cell{\sigma}{\theta'}{\theta}{\sigma'}$. Fixed a decomposition of
 $\cell{\sigma;G}{id_{\underline{1}}}{\theta}{F}$ in terms of basic
 tiles, the proof proceeds by induction on the number of tiles
 associated to the clauses that are considered in the decomposition.
\end{proof}

\begin{corollary}
 For any logic program $\mathcal{P}$, the equivalences
 $\simeq_{\mathcal{P}}$ and $\cong_{\mathcal{P}}$ are congruences with
 respect to conjunction of goals and instantiation of free variables.
\end{corollary}

\subsection{Three goal equivalences via instantiation closures}\label{THREEsec}

 One of the main motivations for the research presented in this paper
 concerns the application of logic programming as a convenient
 computational model for interactive systems. In particular, the
 unification mechanism typical of resolution steps is particularly
 interesting because it differs from the ordinary matching procedures
 of reduction semantics \cite{BB:CAM,Mil:CCS,Mes:CRL}. To some extent,
 mgu's characterizes the minimal amount of dynamic interaction
 with the rest of the system that is needed to evolve. In this section
 we compare other operational alternatives, which are commonly used in
 many concurrent systems and calculi, by means of the equivalences
 they induce on goals.  Each alternative is obtained by slightly
 modifying the operational rule for atomic goals.

 The first model allows for applying only ground instances of the
 clauses (to ground goals only):

\[
 (1) \irule{(H \mathrel{:-} F) \in \mathcal{P}\ \ \
        A = \sigma;H\ \mathrm{ground}\ \ \
        \sigma;F\ \mathrm{ground}} 
        {\mathcal{P}\vvdash A\Rightarrow_{\sigma}
                            \sigma;F}
\]
 
\noindent
 Then, two goals $G_1$ and $G_2$ (not necessarily ground) are
 equivalent, written $G_1 \mathrel{\sim_{(1)}} G_2$ if and only if for
 any ground substitution $\sigma$ on $\mathit{Var}(G_1,G_2)$, whenever
 $\sigma;G_1$ is refuted then also $\sigma;G_2$ is refuted, and vice
 versa. This equivalence is the most widely used for interactive
 systems (closing open systems in all possible ways), since it is the
 coarser ``correct'' equivalence that can be defined according to the
 operational rules. The disadvantage is that to check goal equivalence
 we must instantiate \wrt\ all ground substitutions, \ie, proving goal
 equivalence is in general very expensive.

 The second model allows for applying any instance of the
 clause to any matching instance of the goal:

\[
 (2) \irule{(H \mathrel{:-} F) \in \mathcal{P}\ \ \
        \sigma;A = \sigma;\rho;H} 
        {\mathcal{P}\vvdash A\Rightarrow_{\sigma}
                            \sigma;\rho;F}
\]
 
\noindent
 In this case, two goals $G_1$ and $G_2$ are equivalent, written $G_1
 \mathrel{\sim_{(2)}} G_2$ if and only if whenever $G_1$ can be
 refuted with $\sigma$, then also $G_2$ can be refuted with $\sigma$,
 and vice versa. This equivalence extends the previous one to a
 uniform treatment of open and ground goals, but of course equivalence
 proofs become even more complicated and inefficient.

 The third model is the ordinary one, where the substitution $\sigma$
 in (2) must be the mgu between $A$ and $\rho;H$. Hence, two goals
 $G_1$ and $G_2$ are equivalent, written $G_1 \mathrel{\sim_{(3)}}
 G_2$ if and only if they have the same set of computed answer
 substitutions (\ie, $\mathrel{\sim_{(3)}}$ is the equivalence
 $\mathrel{\simeq_{\mathcal{P}}}$ discussed in Section~\ref{GCsec}).
 This equivalence is very convenient, because it makes the transition
 system finitely branching (as opposed to (1) and (2)) and therefore
 facilitates equivalence proofs.

 If we work with an infinite set of function symbols, it can be easily
 verified that $\mathrel{\sim_{(1)}}$ and $\mathrel{\sim_{(2)}}$
 define exactly the same equivalence classes. The inclusion of
 $\mathrel{\sim_{(2)}}$ into $\mathrel{\sim_{(1)}}$ is obvious,
 because ground substitutions are just a particular case of generic
 substitutions. The converse holds because the existence of a
 refutation with ground substitution $\sigma;\psi$, where $\psi$
 contains function symbols not appearing in the program, implies the
 existence of another refutation with non-ground substitution
 $\sigma$. Therefore, the equivalence over ground substitutions
 ($\mathrel{\sim_{(1)}}$) together with the assumption of an infinite
 set of function symbols imply the equivalence over non-ground
 substitutions ($\mathrel{\sim_{(2)}}$).

 The equivalence $\mathrel{\sim_{(3)}}$ is instead stricter than the
 other two. Again, the inclusion of $\mathrel{\sim_{(3)}}$ in
 $\mathrel{\sim_{(1)}}$ is obvious, while it is easy to find an
 example of a logic program $\mathcal{P}$ where two goals have
 different sets of computed answer substitutions but have the same
 sets of ground refutations. Just consider a logic program with the
 following three facts:

\begin{alltt}
 p(X).
 p(a).
 q(X).
\end{alltt}

\noindent
 If we take the goals $p(X)$ and $q(X)$, it is immediate to see that $p(X)
 \mathrel{\sim_{(1)}} q(X)$. However, the set of computed answer
 substitutions of $p(X)$ is $\{\varepsilon, [\sbtn{a}{X}] \}$, while
 for $q(X)$ we just have $\{\varepsilon\}$ (with $\varepsilon$
 denoting the empty substitution).

\subsection{Concurrency and causality}\label{DTsec}

 If we look at the  system $\mathcal{R}_{\mathcal{P}}$ from a
 concurrent viewpoint then atomic goals can be regarded as distributed
 components that can evolve separately and where variable sharing
 provides the means to exchange information between
 components. According to this perspective,  \eg, for
 backtracking, it is essential to keep
 track of the causal dependencies among components.

 To accomplish this view we slightly modify the associated tile system
 for defining more concrete observations on the causal dependencies
 among replaced and inserted goals. To this aim, to each clause
\[
 c \mathrel{\equiv} p(t_1,\ldots,t_k)\ \mathrel{:-}\
                    q_1(\vec{s}_1),\ldots,q_m(\vec{s}_m)
\]
\noindent
 in the logic program, we associate an operator
 $\dfun{c}{\mathsf{p}^m}{\mathsf{p}}$ in the signature of
 observations. Then, since we want to be aware of the components
 distributed in the system, we do not consider the operator $\wedge$
 and associate to each clause $C$ the tile

\[
 C_c =
 \cell{p}
      {c}
      {\langle t_1,\ldots,t_k\rangle}
      {\langle q_1(\vec{s}_1),\ldots,q_m(\vec{s}_m)\rangle}.
\]

\noindent
 Note that by using the
 trigger $c$ we can now observe that the initial configuration $p$
 generates $m$ new components that causally depends on it.

 For the rest, we add as before the pullback tiles that provide the
 coordination mechanism about local instantiations.  In the new
 setting, equivalent computations from the point of view of
 concurrency and coordination are identified, whereas computations
 that return the same computed answer substitution but that employ
 different concurrent reduction strategies are distinguished. This
 also allows one to observe causal dependencies among resolution
 steps, since the triggers of refutation tiles describe the `concurrent
 strategy' employed for achieving the result.

 Of course, the notion of refutation tile slightly changes according
 to the above modification: A refutation tile is an entailed tile of
 the form $\cell{G}{s}{\theta}{\Box^m\otimes !_{\underline{n}}}$, \ie,
 empty goals become \emph{nil} processes distributed around system
 locations.

 The trigger of the refutation tile for a generic goal is a tuple of
 terms (without shared variables), \ie, it corresponds to an ordered
 forest of (ordered) trees, whose nodes are labeled with clause
 names. We denote by $\sqsubseteq$ the obvious partial order on nodes
 such that $x \sqsubseteq y$ iff $y$ descends from $x$.  Moreover,
 since the tree is ordered, we have an immediate correspondence
 between each clause instance and the subgoal it was applied to. Then,
 we can characterize the concurrency of the framework by means of the
 following theorem.

\begin{theorem}
 Let $\cell{G}{s}{\theta}{\Box^m\otimes !_{\underline{n}}}$ be a tile
 refutation for the goal $G$, and let $\sqsubseteq$ be the partial
 order (forest) associated to $s$. Moreover, let $\preceq$ be any
 total order that extends $\sqsubseteq$. Then by applying the clauses
 associated to the nodes of the tree in the order specified by
 $\preceq$ we obtain again the computed answer substitution $\theta$.
\end{theorem}

 The proof is based on the compositional properties of pullbacks and
 expresses the `complete concurrency' (from the trigger side, not from
 the effect side) of the framework.

 Since application of clauses that do not depend on each other
 (in $\sqsubseteq$) can be executed in any order by choosing suitable
 total orders $\preceq$, it follows that the order in which they are
 executed is not important. Note however that this depends on the
 fact that the coordination mechanism via pullbacks takes care of the
 side-effects of each clause application.

\section{Tiles and constraints}\label{CLPsec}

 In this section we informally discuss how the tile-based approach can
 be extended to deal with constraint logic programming (\textsc{clp})
 \cite{MS:PCI,JM:CLPS}.  Computational equivalences between the
 ordinary operational semantics of \textsc{clp} and the tile semantics
 we will briefly describe in this section can be established by
 results analogous to Theorem~\ref{reptheo}.  The interest in
 constraint satisfaction problems is centered around a powerful,
 declarative mechanism for knowledge representation, as many
 situations can be conveniently modeled by means of constraints on a
 set of objects or parameters. Therefore, constraint logic programming
 is not a language in itself, but can be more precisely seen as a
 scheme, parametric \wrt\ the kind of constraints that can be
 handled. For example, pure logic programming is to some extent a
 version of \textsc{clp} dealing with term equalities over a Herbrand
 universe. The way in which constraints are combined and simplified is
 delegated to a \emph{constraint solver} for efficiency reasons. Other
 formalisms, like \eg\ \emph{constraint handling rules}
 \cite{Fru:CHR}, allow for modeling solvers by means of suitable
 guarded clauses. Usually, constraint programming languages are
 monotonic (or \emph{non-consuming}), that is constraints are never
 deleted from the constraint store; however, in the literature there
 are some variants which allow for constraint consumption
 \cite{BdBP:POSOS}. Both the monotonic and the consuming behaviors
 can be represented in our framework.

 At the abstract level (see \cite{Sar:CC}), a \emph{constraint system}
 can be seen as a pair $\langle D,\vdash\rangle$, where $D$ is a set of
 \emph{primitive constraints} and $\vdash\subseteq \wp(D) \times D$ is
 the \emph{entailment relation}, relating (finite) sets of primitive
 constraints to entailed primitive constraints. Relation $\vdash$ must
 satisfy

\begin{itemize}
\item
 $\mathbf{C}\cup\{\mathbf{c}\} \vdash \mathbf{c}$ (reflexivity);
\item
 if $\mathbf{C} \vdash \mathbf{c}$ for all $\mathbf{c}\in
 \mathbf{C'}$, and $\mathbf{C'}\vdash \mathbf{c'}$, then
 $\mathbf{C}\vdash \mathbf{c'}$ (transitivity).
\end{itemize}

 The set of subsets of $D$ which are closed under entailment is
 denoted by $\abs{D}$, and a \emph{constraint} is just an element of
 $\abs{D}$. As usual, we assume that a set $\mathit{Con} \subseteq
 \abs{D}$ of \emph{consistent constraints} is given such that:

\begin{itemize}
\item
 if $\mathbf{C}\cup \mathbf{C'} \in \mathit{Con}$, then
 $\mathbf{C},\mathbf{C'}\in\mathit{Con}$;
\item
 if $\mathbf{c}\in D$ then $\{\mathbf{c}\}\in\mathit{Con}$;
\item
 if $\mathbf{C}\in\mathit{Con}$ and $\mathbf{C}\vdash \mathbf{c}$, then
 $\mathbf{C}\cup\{\mathbf{c}\}\in \mathit{Con}$.
\end{itemize}

 For our representation, a constraint system can be equivalently seen
 as a category $\mathcal{C}$ whose arrows are constraints and whose
 composition is the conjunction of constraints, in such a way that
 $\mathbf{C};\mathbf{c} = \mathbf{C}$ iff $\mathbf{C}\vdash
 \mathbf{c}$. We also assume a distinguished arrow $\mathbf{ff}$
 exists such that $\mathbf{C} = \mathbf{ff}$ iff $\mathbf{C}\not\in
 \mathit{Con}$.

 In the presence of constraints, goals become pairs $(G,\mathbf{C})$,
 where $G$ is an ordinary conjunctive formula and $\mathbf{C}$ is a
 constraint, while clauses can have the more general form

\[H \mathrel{:-} \mathbf{c_1} | B_1,\ldots,B_n,\mathbf{c_2}\]

\noindent
 where $\mathbf{c_1}$ is a \emph{guard} for the application of the
 clause (similar to the $\mathit{ask}$ operation of \emph{concurrent
 constraint programming} (\textsc{ccp}) \cite{Sar:CC}), and
 $\mathbf{c_2}$ is the constraint to be added to the store after the
 application of the clause (similar to the $\mathit{tell}$ operation
 of \textsc{ccp}). In the ordinary interpretation, the meaning is
 that the clause can be applied only if the constraint component of
 the current goal entails $\mathbf{c_1}$ and that, after the
 resolution step, the constraint $\mathbf{c_2}$ is added to the
 current state, provided that the resulting constraint is consistent,
 \ie, the resolution can be applied only if $\mathbf{C}\vdash
 \mathbf{c_1}$ and $\mathbf{C}\cup \{\mathbf{c_2}\} \in \mathit{Con}$.
 Although usually \textsc{clp} languages do not have guards in their
 syntax but just constraints in the bodies of the clauses (which
 correspond to the tell constraint $\mathbf{c_2}$ above), we decide to
 consider this more general kind of clauses in order to model also
 concurrent formalisms such as \textsc{ccp} and constraint rewriting
 formalisms such as \textsc{chr}.

 Now, we face several alternatives for describing the interaction
 between ask/tell and the current store, where each alternative
 corresponds to a different set of auxiliary tiles associated with the
 constraint system $\mathcal{C}$.

 For example, likewise pure logic programming, we can take the
 pullback squares in $\mathcal{C}$ (if any), or more generally, we can
 consider the \emph{relative pullbacks}, dualizing the approach of
 Leifer and Milner based on relative pushouts \cite{LM:DBCRS}. In this
 case, given a constraint $\mathbf{C}$ (the current constraint store)
 and another constraint $\mathbf{c_2}$ (the tell part of the clause),
 we have coordination tiles of the form
 $\cell{\mathbf{C}}{\mathbf{c_2}}{\mathbf{c}}{\mathbf{C'}}$ (with the
 condition that $\mathbf{c};\mathbf{C} \neq \mathbf{ff}$) expressing
 that $\mathbf{C'}$ is the minimal constraint to be added to
 $\mathbf{c_2}$ for entailing (all constraints in) $\mathbf{C}$, and
 that $\mathbf{c}$ is the minimal constraint to be added to
 $\mathbf{C}$ for entailing $\mathbf{c_2}$. Therefore, tiles of this
 kind check the consistency of $\mathbf{c_2}$ with $\mathbf{C}$ and
 return the additional amount of information gained by joining the two
 constraints, an operation which is suitable for interpreting tell
 constraints. The guard $\mathbf{c_1}$ should be considered as part of
 the initial configuration of the tile associated with the clause, so
 that the clause can be applied only if the current store $\mathbf{C}$
 can be decomposed as $\mathbf{C_1};\mathbf{c_1}$, while
 $\mathbf{c_2}$ is an effect of the tile (to be coordinated with the
 current state). Since the ask and tell operations are not consuming,
 the join of $\mathbf{c_1}$ and $\mathbf{c_2}$ must be inserted also
 in the final configuration. The resulting tile associated with the
 clause is thus
 $\cell{(P,\mathbf{c_1})}{id}{(t,\mathbf{c_2})}{(B_1,\ldots,B_n,\mathbf{c_2};\mathbf{c_1})}$,
 with $P$ the predicate symbol in $H$ and $t;P = H$.

 If the category describing the constraint system does not possess the
 pullbacks, we can consider all commuting squares, instead of just
 (relative) pullbacks. This encoding can be applied to a generic
 category $\mathcal{C}$, but usually involves an infinite number of
 possible closures.

 Notice that this way of modeling \textsc{clp} clauses via tiles
 synchronizes $P$ and $\mathbf{c_1}$ and therefore centralizes the
 control, unless $\mathbf{c_1}$ is the empty constraint.  An
 alternative, which solves this problem and gives more emphasis to the
 interaction between subgoals and constraints, is to leave the
 consistency check of the tell operation to the metainterpreter (\eg,
 by discarding all computations that reach an inconsistent store), and
 use the guard $\mathbf{c_1}$ as an effect, to abandon the centralized
 view. Now, the auxiliary tiles for constraints have just the task of
 checking the entailment of the guard in the current store, and
 therefore we can take all squares of the form
 $\cell{\mathbf{C}}{\mathbf{c_1}}{id}{\mathbf{C}}$ such that
 $\mathbf{C};\mathbf{c_1}=\mathbf{C}$ (there is exactly one cell for
 any $\mathbf{C},\mathbf{c_1}$ such that $\mathbf{C}\vdash
 \mathbf{c_1}$). Since $\mathbf{C}$ appears in the final
 configuration, there is no need for reasserting $\mathbf{c_1}$, and
 thus the tile associated with the generic clause is
 $\cell{P}{id}{(t,\mathbf{c_1})}{(B_1,\ldots,B_n,\mathbf{c_2})}$. Note
 that if tiles like $\cell{\mathbf{C}}{\mathbf{c_1}}{id}{\mathbf{C'}}$
 with $\mathbf{C'};\mathbf{c_1}=\mathbf{C}$ were considered instead,
 then the constraint $\mathbf{c_1}$ might also be consumed during the
 entailment check, unless it was reintroduced in the final
 configuration.

 It is worth to notice that in all these proposals to model
 \textsc{clp} via tiles, the tile approach allows us to clearly
 separate the rules of the program from the coordination mechanism,
 which is dependent just on the category $\mathcal{C}$ of constraints
 under consideration, and on the features we want to model.

\section*{Conclusions and future work}

 In this paper we have used tile logic to model the coordination and
 interaction features of logic programming.  Our approach differs from
 that of \cite{CM:ASSTS}, based on structured transition systems, by
 taking into account the computed answer substitutions instead of just
 the correct answer substitutions. In fact, in \cite{CM:ASSTS}, the
 clauses are seen as rewrite rules that can be further instantiated in
 all possible ways and the computational model of a program is a
 suitable 2-category (\ie, a special kind of double category, where
 the vertical category is discrete and thus only identities are
 allowed as observations). This means that if there exists a
 refutation for the goal $G$ with computed answer substitution
 $\theta$, then in the 2-category model we can find a refutation for
 $\theta;G$ but not necessarily one for $G$. The main advantages of
 our approach \wrt\ the one in \cite{CM:ASSTS} are a finite branching
 operational semantics and the built-in unification
 mechanism. Moreover, the drawback of using 2-categories instead of
 double categories is that the dynamic creation of fresh variables
 cannot be modeled, \ie, the variables to be used must all be present
 at the beginning of the computation.

  As noted in the Introduction, the usage of tiles emphasizes the
 duality of instantiation and contextualization of goals, allowing for
 a uniform treatment of both. In particular, while instantiation plays
 a fundamental role since it can affect the behavior of the goal,
 here contextualization (\ie, conjunction with disjoint goals) does
 not increase the distinguishing power of the semantics, and therefore
 transitions labeled with external contexts can be avoided in the
 model. In fact, each atomic goal can be studied in isolation from
 other goals and the tiles for putting a goal in any possible
 conjunction are not necessary. The reason for this absence is that
 goals cannot be conjoined in Horn clauses' heads, whereas if
 multi-headed clauses were considered, then, in general, abstract
 semantics would not turn out to be a congruence unless transitions
 for external contexts are added.  Indeed, we believe that our
 approach can be extended to other frameworks where multi-head clauses
 are allowed (see for example the \emph{generalized Horn clauses}
 of \cite{FLP:GHC} and the \textsc{chr} formalism
 \cite{Fru:CHR}), giving us the key for dealing with
 contextualization features --- dually to the instantiation via
 pullbacks considered here.

 We have also sketched some ideas for extending our approach to handle
 constraints. In this case, the constraint system and the logic
 program are modeled by two separate sets of tiles, and we have shown
 how to handle both ask and tell constraints.

 Exploiting the built-in synchronization features of tile logic, we
 are confident that our framework can be naturally extended to deal
 also with sequentialized commits (\ie, goals of the form
 $G_1;G_2;\cdots;G_k$ where the possibly non-atomic subgoals $G_i$ must be
 resolved in the order given by their indices $i$, giving the
 possibility to the user of specifying more efficient resolution
 strategies). Moreover, the higher-order version of tile logic
 presented in \cite{BM:CCDC} may find application to the modeling of
 higher-order logic programming (\eg, \emph{lambda prolog})
 \cite{Mil:LPILL,MN:HOLP}.

 Finally, let us mention that the abstractness of the unification via
 pullbacks makes the tile approach suitable for considering
 unification in equational theories rather than in term algebras. For
 example, this would allow to develop a computational model for
 rewriting logic (and hence for reduction systems on processes up to
 structural congruence) based on unification.

\section*{Acknowledgements}

 We would like to thank Paolo Baldan and Fabio Gadducci for their
 comments on a preliminary version of this paper.  We are also
 grateful to the anonymous referees for their helpful suggestions and
 comments, which allowed us to improve the presentation of the
 material.

 This research has been supported by
  CNR Integrated Project
      \emph{Progettazione e Verifica di Sistemi Eterogenei};
  by Esprit WG
      \emph{CONFER2} and \emph{COORDINA};
  and by MURST project
      \emph{TOSCA}.

\end{document}